\documentclass[twocolumn,tighten]{aastex62}

\usepackage{graphicx}
\usepackage{ulem}
\usepackage{amsmath}
\usepackage{wrapfig}

\def\apjl{{ApJL}}
\def\apjs{{ ApJS}}

\def\aap{{ A\&A}}

\def\mr{\mathrm}

\def\d{\mr{d}}

\def\t{\widetilde}
\def\mc{\mathcal}
\def\msun{M_{\rm \odot}}
\def\eps{\mc{E}}
\def\feps{f_{\epsilon}}

\def\rt{r_{\rm TDE}}
\def\rp{r_{\rm p}}
\def\rg{r_{\rm g}}
\def\rc{r_{\rm c}}
\def\rd{r_{\rm d}}
\def\rtr{r_{\rm tr}}

\def\risco{r_{\rm isco}}
\def\vw{v_{\rm w}}

\begin{document}
\shorttitle{BH--star TDE\lowercase{s}}
\shortauthors{Kremer et al.}

\title{Tidal Disruptions of Stars by Black Hole Remnants in Dense Star Clusters}
%% Authors with the same affiliation can be grouped in a single
%% \author and \affil call.
\author[0000-0002-4086-3180]{Kyle Kremer}
\affil{ Department of Physics \& Astronomy, Northwestern University, Evanston, IL 60202, USA}
\affil{ Center for Interdisciplinary Exploration \& Research in Astrophysics (CIERA), Evanston, IL 60202, USA}
\author[0000-0002-1568-7461]{Wenbin Lu}
\affil{TAPIR, Walter Burke Institute for Theoretical Physics, Mail Code 350-17, Caltech, Pasadena, CA 91125, USA}
\author[0000-0003-4175-8881]{Carl L. Rodriguez}
\affil{Pappalardo Fellow; MIT-Kavli Institute for Astrophysics and Space Research, Cambridge, MA 02139, USA}
\author{Mitchell Lachat}
\affil{Department of Physics, Allegheny College, Meadville, Pennsylvania 16335, USA}
\affil{ Center for Interdisciplinary Exploration \& Research in Astrophysics (CIERA), Evanston, IL 60202, USA}
\author[0000-0002-7132-418X]{Frederic A. Rasio}
\affil{ Department of Physics \& Astronomy, Northwestern University, Evanston, IL 60202, USA}
\affil{ Center for Interdisciplinary Exploration \& Research in Astrophysics (CIERA), Evanston, IL 60202, USA}

\begin{abstract}
In a dense stellar environment, such as the core of a globular cluster (GC), dynamical interactions with black holes (BHs) are expected to lead to a variety of astrophysical transients. Here we explore tidal disruption events (TDEs) of stars by stellar-mass BHs through collisions and close encounters. Using state-of-the-art cluster simulations, we show that these TDEs occur at significant rates throughout the evolution of typical GCs and we study how their relative rates relate to cluster parameters such as mass and size. By incorporating a realistic cosmological model of GC formation, we predict a BH -- main-sequence-star TDE rate of approximately $3\,\rm{Gpc}^{-3}\,\rm{yr}^{-1}$ in the local universe ($z<0.1$) and a cosmological rate that peaks at roughly $25\,\rm{Gpc}^{-3}\,\rm{yr}^{-1}$ for redshift 3. Furthermore, we show that the ejected mass associated with these TDEs could produce optical transients of luminosity $\sim 10^{41} - 10^{44}\rm\,erg\,s^{-1}$ with timescales of about a day to a month. These should be readily detectable by optical transient surveys such as the Zwicky Transient Facility. Finally, we comment briefly on BH -- giant encounters and discuss how these events may contribute to the formation of BH -- white-dwarf binaries.

%Finally, we comment briefly on the binary BH (BBH) merger rate and note that core-collapsed clusters (which to date have been ignored in similar studies) yield an increased number of BBH mergers. This suggests that previous predictions of the BBH merger rate from GCs may be low by a small factor.

\end{abstract}

\section{Introduction}
\label{sec:intro}

The high stellar densities at the centers of globular clusters (GCs) make them hotbeds for exotic astrophysical objects, including blue stragglers \citep[e.g.,][]{Sandage1953,Ferraro1995,Piotto2003,Geller2011,Leigh2013,Chatterjee2013b}, millisecond pulsars \citep[e.g.,][]{Lyne1987,Sigurdsson1995,Ransom2008,Ivanova2008,Fragione2018a,Ye2018}, and low-mass X-ray binaries \citep[e.g.,][]{Clark1975,Verbunt1984,Heinke2005,Ivanova2013,Giesler2018,Kremer2018a}. 
%FRED: need more refs (balance older and newer, ours and others', for all 3 things)
The influence of dynamics is particularly pronounced for stellar-mass black holes (BHs), which preferentially reside in the highest-density central regions of their host clusters, as a consequence of mass segregation. Recent work has shown that GCs efficiently produce merging binary BHs (BBHs) through series of binary-mediated dynamical encounters \citep[e.g.,][]{Moody2009,Banerjee2010,Bae2014,Ziosi2014,Rodriguez2015a,Rodriguez2016a,Askar2017, Banerjee2017,Giesler2018,Hong2018,Fragione2018b,Samsing2018a,Rodriguez2018b}. 
%FRED: careful with these refs... 2000 is not "recent" so drop SPZ&MM; I'd list all relevant papers from the last 5 years only, but make sure not to miss relevant works from outside our group...
Thus, as LIGO and Virgo have begun to unveil the gravitational-wave universe through detections of merging BBHs \citep[e.g.,][]{Abbott2016a,Abbott2016c,Abbott2016b,Abbott2016d,Abbott2017,Abbott2018b}, the topic of binary BH formation in GCs has risen in importance.

In the past two decades, our understanding of stellar-mass BH populations in GCs has evolved significantly. From an observational perspective, several BH candidates have been identified in GCs through both X-ray/radio \citep[e.g.,][]{Maccarone2007,Strader2012} and radial-velocity \citep{Giesers2018} observations. Recent computational work has supported these observational findings. Current models of GCs show that they can retain large populations of BHs, even at late times ($t\gtrsim 10$ Gyr) and the BHs readily mix with luminous stellar populations forming BH -- non-BH binaries. Furthermore, it is now understood that the BHs can play a dominant role in determining the large-scale structural properties of their host cluster \citep[e.g.,][]{Merritt2004,Mackey2007,Mackey2008,Peuten2016,Wang2016,ArcaSedda2018,Kremer2018b,Zocchi2019,Kremer2019a}.

In a GC, stars will frequently pass through small-$N$ (most commonly, three- or four-body) resonant encounters that lead to a greatly increased rate of close passages between objects relative to direct single--single interactions \citep[e.g.,][]{Bacon1996, Fregeau2004}. If these small-$N$ encounters involve BHs and other stars, sufficiently close passages can lead to tidal disruption events (TDEs) of stars by the BHs. Previous work has shown that these events may serve as a dynamical formation channels for a variety of sources observed in GCs. For instance, \citet{Ivanova2005} showed that direct physical collisions between neutron stars (NSs) and giant stars can lead to bright, ultracompact X-ray binaries, similar to those observed in Milky Way GCs, in which the NS is accreting from a white-dwarf (WD) companion. Similarly, \citet{Ivanova2010} showed that collisions between BHs and giants may result in accreting BH -- WD binaries, possibly similar to the X-ray/radio source observed in NGC~4472 \citep{Maccarone2007}. More recently, \citet{Ivanova2017} showed that (grazing) tidal captures of subgiants by stellar-mass BHs may lead to BH -- sub-subgiant binaries, similar to the BH candidate observed in M10 \citep{Shishkovsky2018}. 

Additionally, TDEs involving stellar-mass BHs may produce interesting electromagnetic transient signatures. \citet{Perets2016} showed that disruptions of stars by stellar-mass BHs and NSs, which they called ``micro-TDEs," may give rise to bright, energetic, long X-ray/gamma-ray flares, which could possibly resemble ultra-long gamma-ray bursts. Recently, \citet{Lopez2018} demonstrated that when a star is disrupted by a BBH, the BH spins may be significantly modified if a significant fraction of stellar material is accreted. \citet{Samsing2019} showed that for disruptions by BBHs, observations of breaks in the light curve may be used to determine the BBH orbital period, and they proposed to link this inferred orbital period to the BBH merger rate from clusters, thus connecting tidal disruption and gravitational-wave signals. Additionally, \citet{Fragione2019} recently showed that secular effects relevant in hierarchical triple systems may also lead to BH--star TDEs that occur at significant rates.

In this analysis, we use our H\'{e}non-type Monte Carlo code \texttt{CMC} to explore the rates of TDEs by stellar-mass BHs in GCs, and we then estimate the electromagnetic signatures expected from these events. Our paper is organized as follows. In Section~\ref{sec:method} we summarize the methods used to model the long-term evolution of GCs, including our treatment of BH TDEs. In Section~\ref{sec:results} we discuss the total number of BH -- main-sequence-star (MS) TDEs identified in our models and discuss the dependence of these events on cluster properties, in particular the initial cluster size and mass. In Section~\ref{sec:electromagnetic} we discuss the expected electromagnetic signatures of such events and in Section~\ref{sec:rates} we give our predictions for the event rate. We briefly discuss BH--giant encounters in Section~\ref{sec:giants}. Finally, we discuss our results and conclude in Section~\ref{sec:discussion}.

\section{Globular cluster models}
\label{sec:method}

\begingroup
\renewcommand{\arraystretch}{1.6}
\begin{deluxetable*}{c|c|c|c||c|c|c|cccc|c|c}
\tabletypesize{\scriptsize}
\tablewidth{0pt}
\tablecaption{Initial and final cluster properties for all models\label{table:models}}
\tablehead{
	\colhead{Model} &
    \colhead{$N$} &
    \colhead{Z} &
    \colhead{$r_v$}&
    \colhead{$M_{\rm{tot}}$}&
    \colhead{$N_{\rm{BH}}$}&
    \colhead{$N_{\rm{BBH,\,merger}}$}&
    \multicolumn{4}{c}{BH--MS TDEs}&
    \colhead{BH--Giant TDEs} &
    \colhead{BH--WD TDEs}\\
    \colhead{} &
    \colhead{($\times10^5$)} &
    \colhead{($ \rm{Z}_{\odot}$)} &
    \colhead{($\rm{pc}$)}&
    \colhead{($\times10^5\,M_{\odot}$)}&
    \colhead{}&
    \colhead{}&
    \colhead{s--s}&
    \colhead{b--s}&
    \colhead{b--b}&
    \colhead{total}&
    \colhead{}&
    \colhead{}\\
    \colhead{Total}
}
\startdata
1 & 2 & 0.01 & 1 & 0.5 & 1 & 7 & 0 & 2 & 0 & 2 & 1 & 0\\
2 & 5 & 0.01 & 1 & 1.46 & 41 & 42 & 5 & 6 & 2 & 13 & 3 & 0\\
3 & 10 & 0.01 & 1 & 3.09 & 165 & 114 & 23 & 12 & 4 & 39 & 9 & 0\\
4 & 20 & 0.01 & 1 & 6.3 & 634 & 341 & 57 & 49 & 7 & 113 & 29 & 2\\
\hline
5 & 2 & 0.25 & 1 & 0.43 & 1 & 9 & 1 & 2 & 3 & 6 & 6 & 1\\
6 & 5 & 0.25 & 1 & 1.35 & 36 & 48 & 4 & 9 & 3 & 16 & 6 & 0\\
7 & 10 & 0.25 & 1 & 2.84 & 198 & 121 & 17 & 9 & 2 & 28 & 15 & 0\\
8 & 20 & 0.25 & 1 & 5.94 & 608 & 344 & 59 & 53 & 13 & 125 & 29 & 0\\
\hline
9 & 2 & 0.05 & 1 & - & - & 7 & 3 & 0 & 1 & 4 & 3 & 0\\
10 & 5 & 0.05 & 1 & 0.87 & 9 & 32 & 7 & 6 & 6 & 19 & 12 & 0\\
11 & 10 & 0.05 & 1 & 2.45 & 89 & 102 & 18 & 23 & 8 & 49 & 18 & 1\\
12 & 20 & 0.05 & 1 & 5.29 & 423 & 331 & 93 & 71 & 10 & 174 & 57 & 1\\
\hline
13 & 2 & 0.01 & 2 & 0.54 & 32 & 4 & 0 & 1 & 0 & 1 & 0 & 0\\
14 & 5 & 0.01 & 2 & 1.54 & 114 & 27 & 3 & 2 & 0 & 5 & 3 & 0\\
15 & 10 & 0.01 & 2 & 3.19 & 416 & 66 & 9 & 8 & 0 & 17 & 2 & 0\\
16 & 20 & 0.01 & 2 & 6.61 & 1192 & 196 & 26 & 21 & 5 & 52 & 9 & 0\\
\hline
17 & 2 & 0.25 & 2 & 0.47 & 23 & 10 & 0 & 0 & 0 & 0 & 0 & 0\\
18 & 5 & 0.25 & 2 & 1.4 & 136 & 20 & 3 & 2 & 1 & 6 & 1 & 0\\
19 & 10 & 0.25 & 2 & 2.98 & 415 & 75 & 18 & 5 & 2 & 25 & 3 & 0\\
20 & 20 & 0.25 & 2 & 6.3 & 1133 & 203 & 30 & 11 & 1 & 42 & 8 & 0\\
\hline
21 & 2 & 0.05 & 2 & - & - & 8 & 1 & 1 & 0 & 2 & 1 & 0\\
22 & 5 & 0.05 & 2 & 0.8 & 55 & 24 & 3 & 7 & 0 & 10 & 0 & 0\\
23 & 10 & 0.05 & 2 & 2.48 & 260 & 59 & 8 & 10 & 5 & 23 & 2 & 0\\
24 & 20 & 0.05 & 2 & 5.78 & 871 & 177 & 29 & 23 & 2 & 54 & 12 & 0\\
\hline
$25^{\star}$ & 10 & 0.05 & 1 & 2.82 & 171 & 111 & 67 & 56 & 6 & 129 & 18 & 0\\
$26^{\star}$ & 10 & 0.05 & 2 & 2.97 & 375 & 78 & 22 & 24 & 1 & 47 & 6 & 0\\
\enddata
\tablecomments{Initial and final cluster parameters for the models considered in this study. Columns 2, 3, and 4 show the particle number ($N$), metallicity (Z), and initial virial radius ($r_v$), respectively. Column 5 shows the final cluster mass (at $t=12\,$Gyr) and column 6 shows the total number of BHs retained in the cluster at this time (note that these values are null for models 9 and 21, which have both dissolved by 12 Gyr). Column 7 show the cumulative number of BBH mergers occurring in each model. Columns 8-10 show the total number of BH--MS TDEs that occur through single--single (s--s), binary--single (b--s), and binary--binary (b--b) encounters, respectively and column 11 shows the total number of BH--MS TDEs in each model. Columns 12 and 13 show the total number of BH--giant and BW--WD TDEs which occur in each model. The final two rows (models 25 and 26, marked with asterisks) show models for which we place a less stringent requirement on the minimum pericenter distance for a BH--MS TDE to occur (see Section \ref{sec:method} for more detail).}
\end{deluxetable*}
\endgroup

\texttt{CMC} (for Cluster Monte Carlo) is a rigorously tested, H\'{e}non-type Monte Carlo code that computes the long-term evolution of GCs with realistic initial properties and containing up to several million stars \citep{Henon1971a,Henon1971b,Joshi2000,Joshi2001,Fregeau2003,Pattabiraman2013,Chatterjee2010,Chatterjee2013,Rodriguez2015a}. \texttt{CMC} incorporates stellar and binary evolution using the evolution codes \texttt{SSE} and \texttt{BSE} \citep{Hurley2000,Hurley2002}, modernized to reflect our most up-do-date understanding of compact object formation, including prescriptions for natal kicks, mass-dependent fallback, and (pulsational) pair-instability supernovae \citep{Fryer2001,Belczynski2002,Hobbs2005,Morscher2015}. We also integrate all small-$N$ resonant dynamical encounters (including binary--single and binary--binary encounters) using the \texttt{fewbody} package \citep{Fregeau2004,Fregeau2007} which has been updated to incorporate gravitational radiation reaction for all encounters involving BHs \citep[for more information, see][]{Rodriguez2018b, Rodriguez2018a}.

We use the same prescription to treat pulsational-pair instabilities and pair-instability supernovae as \citet{Rodriguez2018a}, which follows \citet{Belczynski2016b}. Briefly, we assume that any star with a pre-explosion helium core between 45 and 65 $M_{\odot}$ will undergo pulsations that eject a significant amount of mass, until the final product is at most 45 $M_{\odot}$. Assuming that $10\%$ of the mass is lost during the conversion from baryonic to gravitational mass at the time of collapse, this leaves a BH of mass $40.5\,M_{\odot}$. Finally, we assume that stars with helium core masses in excess of $65\,M_{\odot}$ are completely destroyed by a pair-instability supernova (no remnant is formed). Of course, there are many uncertainties associated with these numbers. In particular, some studies, \citep[e.g.,][]{Woosley2016, Spera2017} have predicted that the lower boundary of this mass gap may be as high as $50\,M_{\odot}$.

For NS formation and evolution, we have updated the \texttt{SSE} and \texttt{BSE} codes to include changes to the magnetic field and spin-period evolution, as well as the natal kick prescriptions for NSs formed in ECSNe \citep{Kiel2008,KielHurley2009}. See \citet{Ye2018} for further detail on the treatment of NSs and radio pulsars in \texttt{CMC}.

In a GC, the stellar dynamics in the high-density core can frequently lead to close encounters between stars. For encounters of two single stars with masses $M_1$ and $M_2$, the cross-section for an encounter with pericenter distance $r_p$ is given by

%FRED: please check carefully; I tried to fix the text here, which made no sense.
\begin{equation}
\label{eq:Sigma_strong}
\Sigma_{\rm{enc}}(r_p) =\pi r_p^2 \Big[1+\frac{2G(M_1+M_2)}{r_p \sigma_v^2} \Big]    
\end{equation}
where $\sigma_v$ is the local velocity dispersion \citep[see, e.g.][]{Fregeau2007}. Here, we are interested in the particular case of close encounters between BHs and MS stars. For sufficiently close encounters, the tides raised on the MS star may dissipate enough energy to bind the two objects or disrupt the structure of MS star. Following \citet[]{Fabian1975}, we define the characteristic radius for tidal interaction as
\begin{equation}
\label{eq:r_TD}
r_{\rm{T}} = f_p\Big( \frac{M_{\rm{BH}}}{M_{\star}}\Big)^{1/3}\,R_{\star},
\end{equation}
where $M_{\rm{BH}}$ is the BH mass, $M_{\star}$ and $R_{\star}$ are the stellar mass and radius, respectively, and $f_p$ is a dimensionless parameter that captures the details of the tidal interaction. In order of increasing $r_p$, close encounters may take the form of physical collisions \citep[e.g.,][]{FryerWoosley1998,Hansen1998,ZhangFryer2001}, tidal disruptions \citep[e.g.,][]{Perets2016}, tidal captures \citep[e.g.,][]{Fabian1975,Ivanova2017}, or more distant tidal encounters that only weakly perturb the star \citep[e.g.,][]{AlexanderKumar2001}. In addition to the dependence on the pericenter distance, the details of a particular close encounter and associated tidal interactions also depend upon the internal structure of the star. For instance, tidal interactions can be quite different for young low-mass MS stars, which have relatively uniform density, compared to evolved massive MS stars which are much more centrally concentrated.

Capturing the subtleties of tidal interactions during close encounters requires detailed hydrodynamic calculations of the events, and is beyond the computational scope of \texttt{CMC}. Therefore, we simply consider two limits for our choice of the maximum pericenter distance, $r_{\rm{TDE}}$, that leads to a TDE. As a conservative lower limit, we take $r_{\rm{TDE}}=R_{\star}$ (models 1-24 in Table \ref{table:models} adopt this assumption) and as an upper limit, we assume $r_{\rm{TDE}}=r_{\rm{T}}$ computed by equation \ref{eq:r_TD}  with $f_p=1$ (models 25-26 adopt this assumption).\footnote{The choice of $f_p=1$ as our upper limit is meant to simply reflect the classical tidal disruption limit assuming a star of uniform density. While this approximation is reasonable for  low-mass MS stars, a different choice for $f_p$ may be more appropriate for evolved massive stars. As this is a first attempt at studying these TDEs in our cluster models, we adopt $f_p=1$ for all stars for simplicity and reserve a more detailed exploration of the dependence of $f_p$ on stellar type for a later study. We do emphasize however, that a stellar-type-dependent $f_p$ is unlikely to change the results presented here significantly.} Henceforth, we use the term TDE as shorthand to denote all close encounters with $r_p<r_{\rm{TDE}}$, computed in one of these two limits. 

Although BH -- MS-star TDEs are the main focus of this study, we comment briefly on NS -- MS-star TDEs in Section \ref{sec:NSTDE}. These are computed in the same way as described above for BHs. In addition to close encounters between MS-stars and BHs (NSs), we also record all close encounters of BHs (NSs) with giants and white dwarfs (WDs). For these events, we consider a maximum pericenter distance of $r_{\rm{TDE}}=R_{\star}$ for all models, where $R_{\star}$ is the stellar radius of the giant or WD. We discuss BH--giant and BH--WD TDEs further in Section \ref{sec:results} and we examine BH--giant encounters specifically in Section \ref{sec:giants}.

For stellar pairs where neither object is a BH or a NS (for example, MS -- MS) \texttt{CMC} includes only physical collisions, treated in the ``sticky-sphere'' approximation, i.e., the two stars are merged assuming conservation of mass and momentum whenever $r_p < R_1 + R_2$, where $R_1$ and $R_2$ are the radii of the two stars \citep[for more detail, see]{Chatterjee2013b}. All stellar radii are computed using the evolutionary tracks of \texttt{SSE} \citep{Hurley2000}.

Close encounters leading to TDEs can occur through both single--single dynamical encounters as well as binary-mediated encounters (binary--single and binary--binary encounters; higher multiples are not included in \texttt{CMC}). For a discussion of how TDEs are computed in \texttt{CMC}, we direct the reader to \citet{Fregeau2007}, which describes the elements of the calculation for both the single--single case and the binary-mediated case in detail. For a more general discussion of the calculation of TDEs and collisions within a Monte Carlo-type dynamics code, we direct the reader to \citet{Chatterjee2010,Goswami2012,Pattabiraman2013} as well as earlier work by \citet{Freitag2002} and \citet{Giersz2001}.

The final outcome of TDEs involving stellar-mass BHs (and NSs) is highly uncertain. Performing detailed calculations of the 3-D hydrodynamics associated with these events is beyond the scope of this study (see \citet{Fixelle2019} for a first attempt). Here, for simplicity, we assume that in the event of a BH--star (or NS--star) TDE, the star is instantaneously destroyed, and no mass is accreted by the BH (NS). Qualitatively, this is because the accretion of even a small amount of mass onto the BH (NS) is expected to easily release enough energy to completely unbind the stellar debris. This process is discussed in more detail in Section \ref{sec:electromagnetic}. However, the possibility that a significant fraction of the disrupted star could be accreted by the BH cannot be entirely ruled out. If indeed a significant portion is accreted, it could affect the cluster's BH population, as discussed further in Section \ref{sec:effects}.

For this analysis, we use the same set of models described in \citet{Rodriguez2018b} (models 1-24), plus two additional models (models 25-26) that adopt an upper limit for tidal disruptions, as described above. A number of initial cluster parameters are fixed throughout, including the initial King concentration parameter ($W_0=5$) and the stellar initial mass function (IMF; we assume an IMF as in \citet{Kroupa2001} with initial masses sampled in the range $0.1-150\,M_{\odot}$). We assume an initial binary fraction of $10\%$ in all models with binary mass ratios drawn from a flat distribution, binary semi-major axes are drawn from a distribution flat in log from near contact to the hard/soft boundary, and initial eccentricities drawn from a thermal distribution. Four initial parameters are varied in this grid of models: the total particle number ($N=2\times 10^5$, $N=5\times 10^5$, $N=10^6$, and $N=2\times 10^6$), the virial radius ($r_v=1$ and 2 pc), the metallicity ($\rm{Z}=0.01$, 0.05, and 0.25 $\rm{Z}_{\odot}$), and the galactocentric distance ($r_{\rm{gc}}=2$, 8, and 20 kpc). We assume that the metallicity and galacocentric distance values are correlated, giving us a $4\times3\times2$ model grid. Table \ref{table:models} lists the initial parameters for all models used in this study, as well as the total number of BHs retained at $t=12\,$Gyr, the cumulative number of BBH mergers, and cumulative number of BH--MS, BH--giant, and BH--WD TDEs in each model. All together, this grid of models approximates the full distribution of cluster masses, sizes, and metallicities observed in the Milky Way.
%FRED: you never mentioned BH--WD before this paragraph!

\section{BH--MS TDE\lowercase{s} in cluster models}
\label{sec:results}

\begin{figure}
\begin{center}
\includegraphics[width=0.5\textwidth]{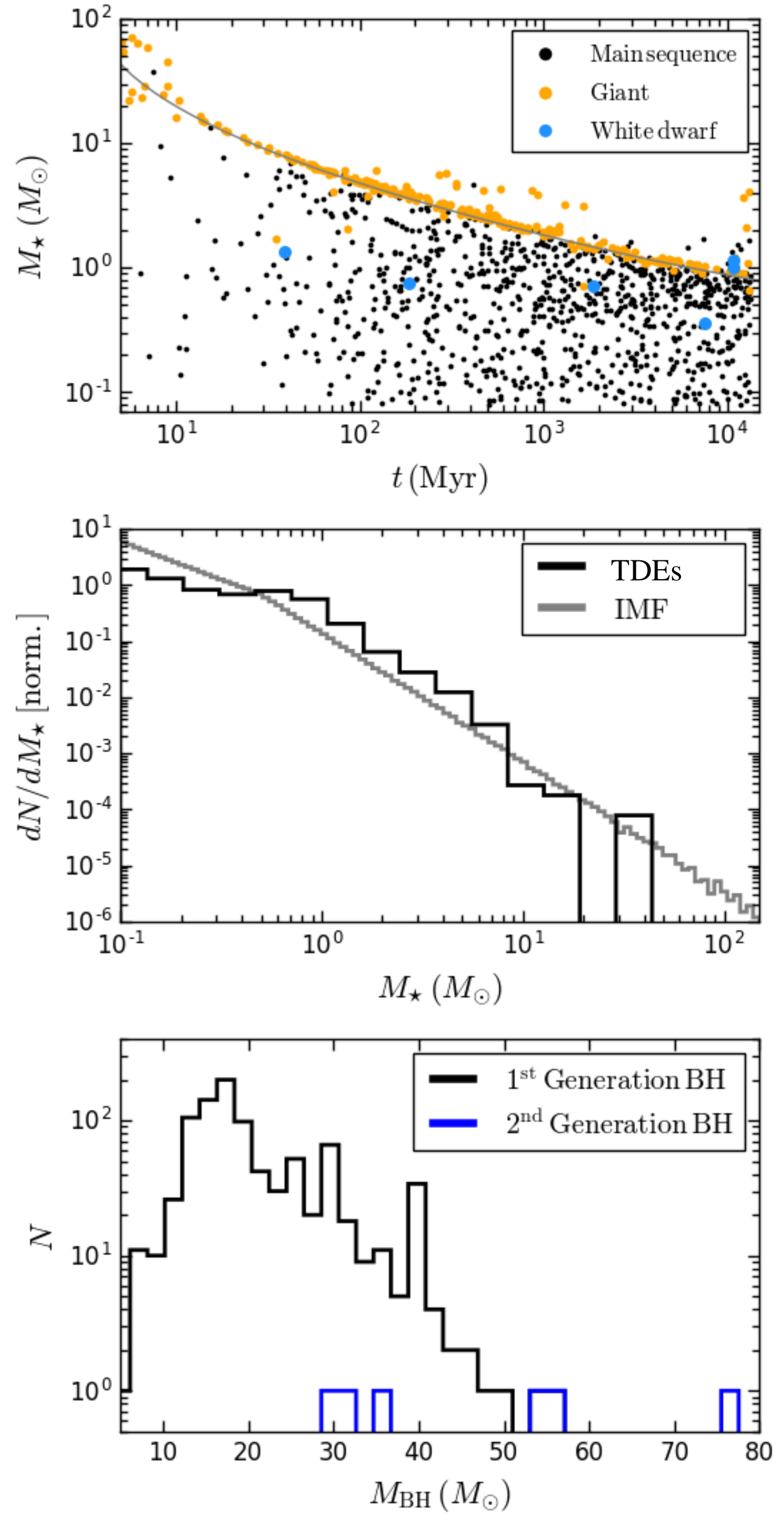}
\caption{\label{fig:scatter} \textit{Top panel:} Stellar mass versus time of TDE for all BH--star TDEs identified in models 1-24. Black, orange, and blue symbols show MS, giant, and WD TDEs, respectively. The solid gray line shows the turnoff mass as a function of time. \textit{Middle panel:} Distribution of stellar mass for all BH--MS TDEs (black) compared to the initial mass function (IMF; gray). The median MS mass is approximately 0.5 $M_{\odot}$. \textit{Bottom panel:} Distribution of BH mass for all BH--MS TDEs. The median BH mass is approximately 18 $M_{\odot}$. The black curve shows first generation BHs and blue shows second generation (BHs that were formed through binary BH mergers earlier in the cluster evolution).}
\end{center}
\end{figure}

As shown in Table \ref{table:models}, we identify 898 total BH--MS TDEs in models 1-24 of this study. Of these, 433, 363, and 102 occur through single--single, binary--single, and binary--binary encounters, respectively. For models 24-25 (for which we increase the cross section for TDEs to obtain an upper limit; see Section \ref{sec:method}), we identify 176 total TDEs, including 89, 80, and 7 from single--single, binary--single, and binary--binary encounters, respectively. Figure \ref{fig:scatter} shows all BH--star TDEs identified in models 1-24. For the top panel, we show the mass of the star involved in each TDE on the vertical axis and we show the TDE time on the horizontal axis. Black points indicate BH--MS TDEs, orange points indicate BH--giant TDEs, and blue points indicate BH--WD TDEs. In total we identify 246 BH--giant TDEs and 6 BH--WD TDEs in models 1-24. We discuss BH--giant TDEs in more detail in Section \ref{sec:giants}. Because BH--WD TDEs occur at such a relatively low rate compared to the other stellar types, we do not consider them further in this analysis.

The gray curve in the top panel of Figure \ref{fig:scatter} marks the turnoff mass as a function of time. As seen in the figure, a handful of BH--MS TDEs involve MS stars with masses lying above the turnoff curve. These MS stars are themselves collision products of MS stars and would be identified as blue stragglers \citep[e.g.,][]{Sandage1953,Ferraro1995,Piotto2003,Chatterjee2013b}.

The middle panel of Figure \ref{fig:scatter} shows the distribution of MS masses for all BH--MS TDEs shown in the top panel (black curve) compared to the initial mass function (IMF; gray curve). As the middle panel shows, the mass distribution of MS stars disrupted by BHs follows the IMF closely. The median MS mass for all BH--MS TDEs is $0.5\,M_{\odot}$.

We show in the bottom panel of Figure \ref{fig:scatter} the distribution of BH masses for all BH--MS TDEs. In black, we show all first generation BHs (BHs that have not already undergone a binary BH merger) and in blue we show all second generation BHs (BHs that were formed from binary BH mergers earlier in the cluster evolution). The peak at $M_{\rm{BH}}=40.5\,M_{\odot}$ in the black curve comes from our treatment of pair instability supernovae (see Section \ref{sec:method}). The handful of first generation BHs with masses slightly above 40.5 $M_{\odot}$ are the result of either stable mass transfer or stellar collisions prior to BH formation (see \citet{Rodriguez2018a} for further detail). In total, only 6 of the roughly 900 total BH--MS TDEs in models 1-24 ($\approx 1\%$) occur with a second generation BH. The median BH mass for all BH--MS TDEs is $18\,M_{\odot}$.

\begin{figure}
\begin{center}
\includegraphics[width=0.5\textwidth]{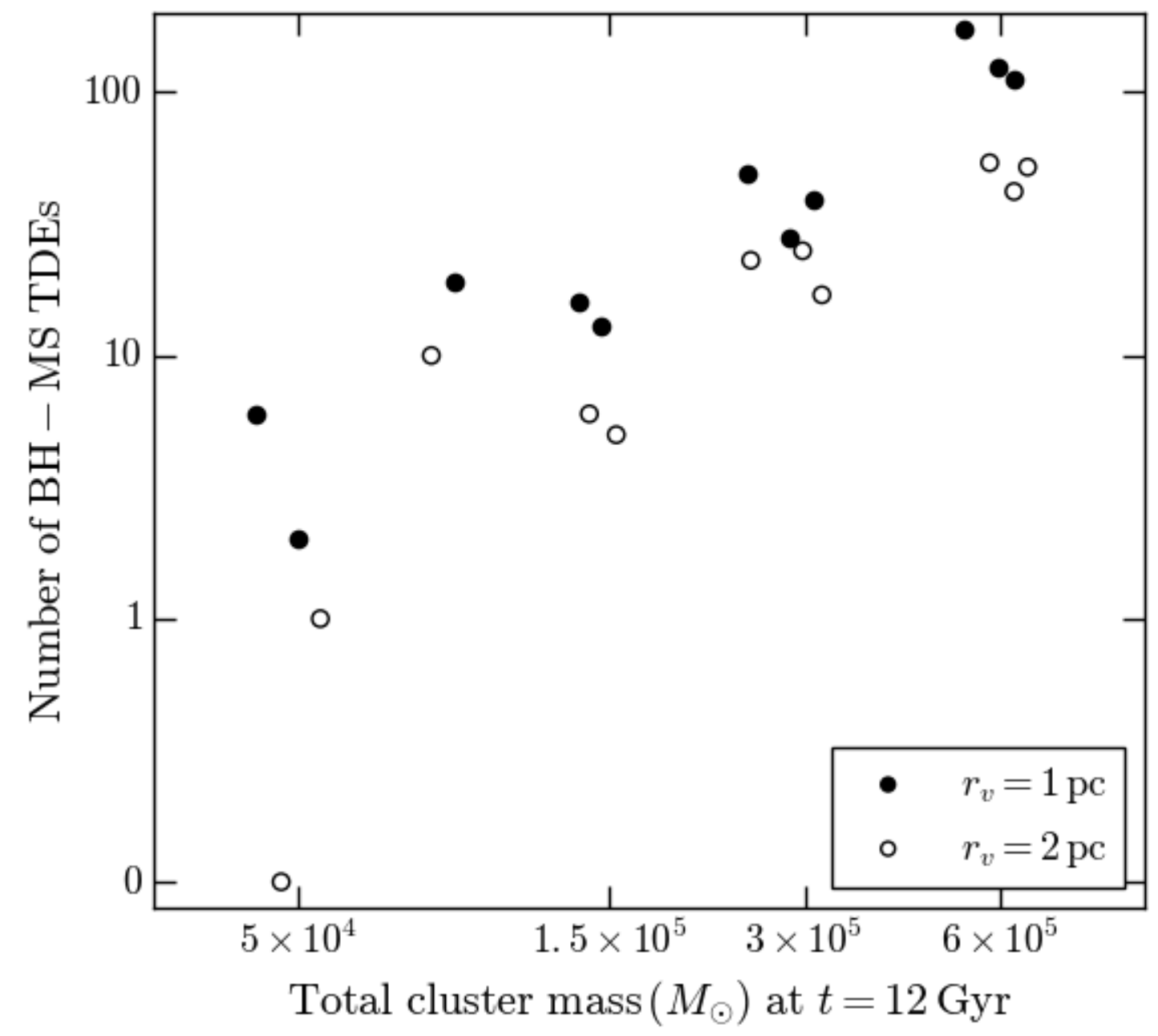}
\caption{\label{fig:models} Total number of BH--MS TDEs in models 1-24 final cluster mass. Filled (open) circles show models with $r_v=1$ (2) pc.}
\end{center}
\end{figure}

In Figure \ref{fig:models} we show how the number of BH--MS TDEs per cluster varies with cluster mass and size (which are specified in our models by initial $N$ and $r_v$, respectively). Here, filled and open circles denote cluster models with $r_v=1\,$pc and $r_v=2\,$ pc, respectively. As the figure shows, the number of TDEs varies proportionally with cluster mass and inversely with cluster size. This is as anticipated. The TDE rate scales with the cluster number density, $n \sim N/r_v^3$. For clusters of similar total $N$, decreasing the size ($r_v$) will increase the density leading to a higher TDE rate. Likewise, for clusters of similar physical size, increasing the number of particles (total mass) will increase the density and again lead to more TDEs. 

For all initial models in this study, we adopt a King concentration parameter of $W_0 =5$ and vary the initial density simply using the initial virial radius. However, in principle, higher values of $W_0$ may also yield an increase in the TDE rate and stellar collision rate. Along these lines, several recent papers \citep[e.g.,][]{PortegiesZwart2004,Freitag2006,Giersz2015,Mapelli2016} have shown that large initial concentrations may even lead to runaway mergers of MS stars in clusters, a possible channel for forming intermediate-mass BHs (IMBHs). An exploration of the effect of initial cluster concentration on the TDE rate is beyond the scope of the present study but is a topic worth consideration in future analyses. We return to this topic of IMBHs briefly in Section \ref{sec:future}.

%These relations can be understood through simple assessment of the expression for the rate of BH--MS collisions (per BH) from single--single encounters

%\begin{equation}
%\label{eq:coll_rate}
%\Gamma_{\rm{coll}} = n_{\star} \Sigma_{\rm{coll}} \sigma_v
%\end{equation}
%Here, $n_{\star}$ is the number density of MS stars, $\Sigma_{\rm{coll}}$ is the cross section for collision (including gravitational focusing), as discussed in Section \ref{sec:method}, $\sigma_v$ is the local velocity dispersion, and $N_{\rm{BH}}$ is the total number of BHs in the cluster.

%Note that $n_{\star} \approx N_{\star} r_v^{-3}$ and assuming virial equilibrium, $\sigma_v \propto (N_{\star}/r_v)^{1/2}$. Also note that in the regime where gravitational focusing dominates, $\Sigma_{\rm{coll}} \propto \sigma_v^{-2}$ (see Equation \ref{eq:Sigma_strong}) which yields $\Sigma_{\rm{coll}} \propto (N_{\star}/r_v)^{-1}$. This allows us to rewrite Equation \ref{eq:coll_rate} as $\Gamma_{\rm{coll}} \propto N_{\star}^{1/2} r_v^{-5/2}$, confirming the relations shown in Figure \ref{fig:models} and Table \ref{table:models}. \textbf{need to make this a little simpler. More concise.}

Finally, comparing models 25 and 26 to models 7 and 19, respectively (which have the same initial conditions), we see that using the upper limit for TDEs described in Section \ref{sec:method}, the number of BH--MS TDEs is increased by a factor of a few (roughly 5 for model 25 versus model 7 and roughly 2 for model 26 versus model 19). This is anticipated: the median BH mass of all disruptions is approximately $18\,M_{\odot}$ and the median MS mass is approximately $0.5\,M_{\odot}$, thus the impact parameter for disruptions is increased, on average, by a factor of $(M_{\rm{BH}}/M_{\star})^{1/3} \approx 3$  compared to our conservative lower limit where we assume $r_{\rm TDE} = R_{\star}$ (see Section \ref{sec:method} for further detail).

\section{Electromagnetic Signatures}
\label{sec:electromagnetic}

%FRED: notations and language in this section need to be updated to become consistent with sec. 2...

In this section we present an analytic prediction for the electromagnetic signature associated with BH--MS TDEs. We start with a broad-brush picture of the disruption process and disk formation from the fallback debris (Section \ref{sec:DiskFormation}). Afterwards, we adopt a simple model for the evolution of super-Eddington accretion disk and its wind (Section \ref{sec:SuperEddington}). The wind carries the radiation generated in the disk (by viscous dissipation) and releases it at larger radii where photons can diffuse away (Section \ref{sec:Radiation}). The final results, the lightcurves and temperature evolution, are shown in Figures \ref{fig:lc_0p5} and \ref{fig:lc_5}. The possibility of jet formation is discussed in Section \ref{sec:GRB}.

\subsection{Tidal disruption and disk formation}
\label{sec:DiskFormation}
Consider a MS star of radius $R_* = r_*R_{\odot}$ and mass $M_*=m_*\msun$, BH mass $M = 10M_1\msun$; the mass ratio is $q = M_*/M$. Adopting the upper limit case discussed in Section  \ref{sec:method} where we assume the maximum pericenter distance that will lead to a TDE, $r_{\rm{TDE}}$, is equal to $r_{\rm{T}}$ as given by equation \ref{eq:r_TD}, we can write
\begin{multline}
  \label{eq:1}
  r_{\rm{TDE}} = R_* q^{-1/3} \simeq (1.0\times10^5\rg)\,
  M_1^{-2/3} m_*^{-1/3} r_*, 
\end{multline}
where $\rg = GM/c^2= 15M_1\rm\,km$ is the gravitational radius of the BH. For stellar-mass BHs with masses up to $\sim 50 \msun$ and MS stars with masses as low as $\sim 0.1 \msun$, equation (\ref{eq:1}) gives $r_{\rm{TDE}}/R_*\sim q^{-1/3}\lesssim 8$ (as compared to $r_{\rm{TDE}}/R_*\gg 1$ for MS-disruptions by supermassive BHs). The initial orbit of the star has pericenter distance $\rp = r_{\rm{TDE}}/\beta$, where $\beta\gtrsim 1$ is the dimensionless impact parameter. Here, we only consider the most frequent cases where $\beta$ is close to unity. Since $r_{\rm{TDE}}$ is comparable to the stellar radius , the star's self-gravity cannot be ignored in the hydrodynamic disruption process. This suppresses the amount of marginally bound debris and hence the late-time fallback rate evolution is steeper than the canonical power-law $t^{-5/3}$ \citep[e.g.][]{2013ApJ...767...25G, Perets2016}.

Instead of modeling the detailed disruption process \citep[see][]{Perets2016}, we focus on the broad-brush picture of disk formation, viscous accretion, and the electromagnetic signals. Roughly speaking, the stellar debris has a spread in specific (kinetic $+$ potential) energy
\begin{equation}
  \label{eq:2}
  \eps \simeq {GM\over r_{\rm{TDE}}} - {GM \over \rt - \eta R_*}
\end{equation}
and specific angular momentum
\begin{equation}
  \label{eq:3}
  \ell \simeq (\rt - \eta R_*) \sqrt{2GM/\rt},
\end{equation}
where the parameter $\eta\in [-1, 1]$ roughly corresponds to fluid elements at different locations inside the star ($\eta=1$ for the part of the star closest to the BH at disruption). The bound debris corresponds to negative specific energy ($0<\eta <1$) and has eccentricity
\begin{equation}
  \label{eq:4}
  e = \sqrt{1 + 2\eps\ell^2/\rg^2c^4} \simeq \sqrt{1 - 4\eta q^{1/3} (1 - \eta q^{1/3})}.
\end{equation}
We define a circularization radius $\rc$ corresponding to a circular Keplerian orbit with the angular momentum in equation (\ref{eq:3}),
\begin{equation}
  \label{eq:5}
  \rc = \ell^2/\rg c^2\simeq 2\rt (1- \eta q^{1/3})^2.
\end{equation}
For stellar mass BHs of mass $M\sim 20\msun$ (roughly the typical BH mass identifed in our cluster models; see Section \ref{sec:results}), we see that $1-e$ is of order unity and hence the orbits of the bound debris are not highly eccentric. This is very different from TDEs by supermassive BHs where we have $1-e\sim (M_*/M)^{1/3}\ll 1$. Therefore, as seen in smoothed-particle hydrodynamics (SPH) calculations \citep[see, e.g.,][]{Fixelle2019}, we expect a nearly circular accretion disk to form quickly after the most tightly bound mass falls back to the pericenter, and the radius of the disk is given by $\rc \sim \rt$ (we ignore the weak dependence on $\eta$).

The viscous accretion timescale at radius $r\sim \rc$ is
\begin{equation}
      \label{eq:6}
  t_{\rm vis} \simeq {(\rc/H)^2 \over \alpha \Omega_{\rm K}} \sim (0.74\mr{\,d})\, (\alpha_{-1})^{-1}
  m_*^{-1/2} r_*^{3/2},
\end{equation}
where $\Omega_{\rm K}(\rc) = \sqrt{GM/\rc^3}$ is
the Keplerian angular frequency,
$\alpha\sim0.1\alpha_{-1}$ is the 
dimensionless viscosity parameter \citep{1973A&A....24..337S}, and we have taken the disk height ratio $H/\rc$ to be 0.5 (appropriate for a super-Eddington thick disk and the uncertainty can be absorbed into $\alpha$). Since the late-time fallback rate drops rapidly on timescale $\Omega_{\rm K}^{-1}$ \citep{Perets2016}, nearly all the bound debris will fall back within the viscous time $t_{\rm vis}(\gg \Omega_{\rm K}^{-1})$ and accumulate near $r\sim \rc$. Thus, we take the disk mass to be $\sim M_*/2$, and then the peak accretion rate is given by
\begin{equation}
  \label{eq:7}
  \dot{M}_{\rm peak} \sim {M_*/2\over t_{\rm vis}}\sim (2.5\times10^2\mr{\,\msun\,yr^{-1}}) {\alpha_{-1} m_*^{3/2}\over r_*^{3/2}}.
\end{equation}
Note that the maximum luminosity of the accretion system is $\sim 0.1\dot{M}_{\rm peak}c^2\sim 1.4\times10^{48} (\alpha/0.1)\mr{\,erg\,s^{-1}}$, which can only be achieved under conditions for extremely efficient energy release (e.g. in the form of relativistic jets, see \S 4.4). Defining the Eddington accretion rate as $\dot{M}_{\rm
  Edd} \equiv L_{\rm Edd}/c^2 = 2.6\times10^{-8}M_1 \mr{\,\msun\,yr^{-1}}$ ($L_{\rm Edd}$
being the Eddington luminosity), we have $\dot{M}_{\rm peak}\rg/\dot{M}_{\rm Edd}\rc \sim 10^5\alpha_{-1} \gg 1$. Thus, the viscous heating rate $\sim GM\dot{M}_{\rm peak}/\rc$ exceeds the Eddington luminosity by about five orders of magnitude and hence the disk is indeed geometrically thick near $r\sim \rc$.

\subsection{Super-Eddington accretion disk}
\label{sec:SuperEddington}
\begin{figure}
  \centering
\includegraphics[width = 0.48\textwidth,
  height=0.38\textheight]{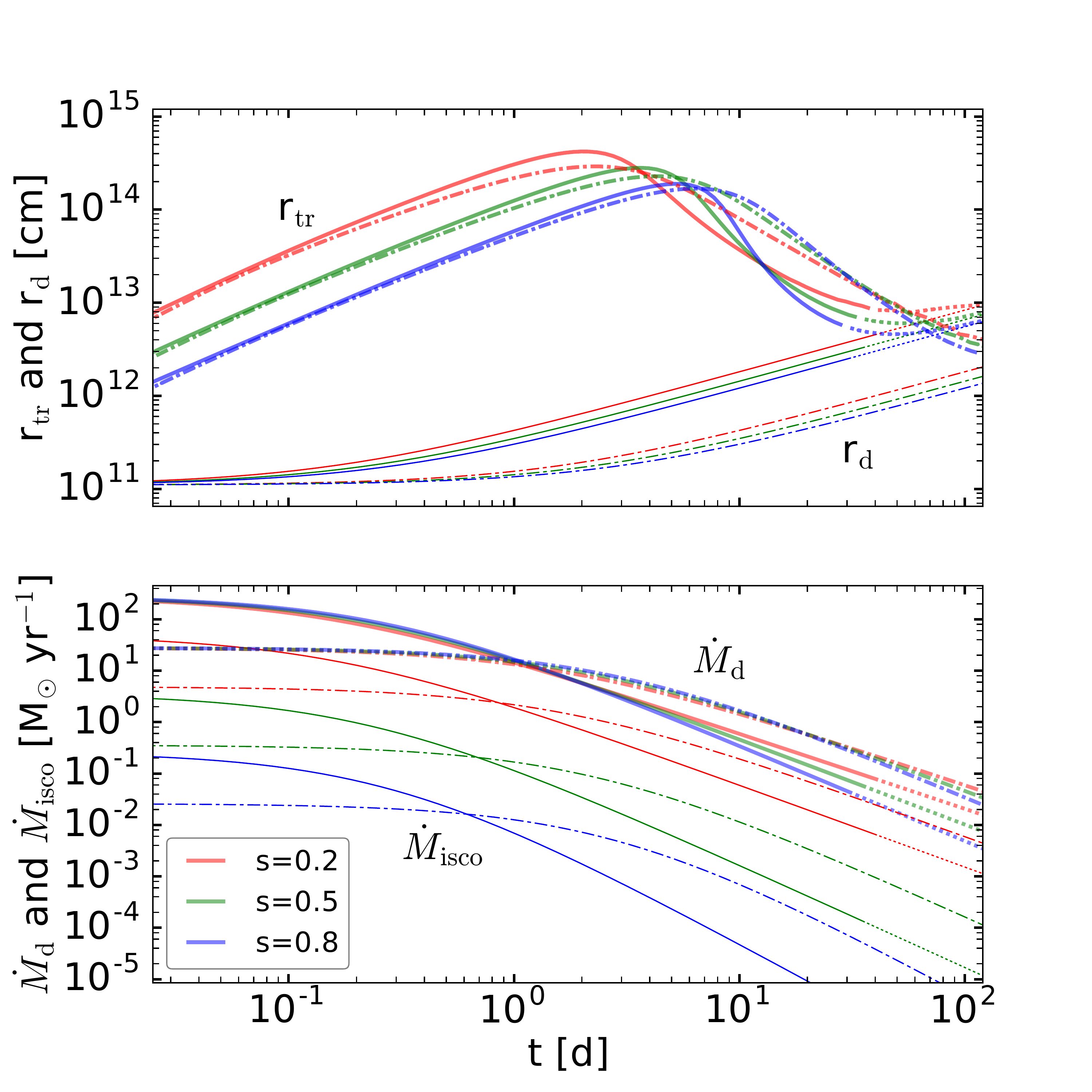}
\caption{\textit{Top panel}: The time evolution of the photon-trapping radius $\rtr$ (equation \ref{eq:r_trap}, thick curves) and the disk radius $\rd$ (thin curves). \textit{Bottom panel}: The accretion rates at the outer radius $\dot{M}_{\rm d}$ (thick curves) and at the ISCO $\dot{M}_{\rm isco} = \dot{M}_{\rm d}(\risco/\rd)^s$ (thin curves). 
For both panels, three cases of different mass flux power-law indexes $s$ (equation \ref{eq:8}) are shown in red ($s=0.2$), green ($0.5$), and blue ($0.8$). Two cases of different viscosity parameters $\alpha$ are shown in solid curves ($\alpha=0.1$) and dash-dotted curves ($\alpha=0.01$).
We fix the efficiency of escaping power $\feps=0.5$ (equation \ref{eq:11}), ISCO radius $\risco=6\rg$, BH mass $M=20\msun$, and stellar mass $M_*=0.5\msun$. The solid (and dash-dotted) segment of each curve corresponds to when the accretion disk stays thick, and the dotted segment at late time indicates that the disk may collapse into a thin one (so our assumption of $H/r\sim 1$ becomes invalid).
}\label{fig:disk_evolution_0p5}
\end{figure}

\begin{figure}
  \centering
\includegraphics[width = 0.48\textwidth,
  height=0.38\textheight]{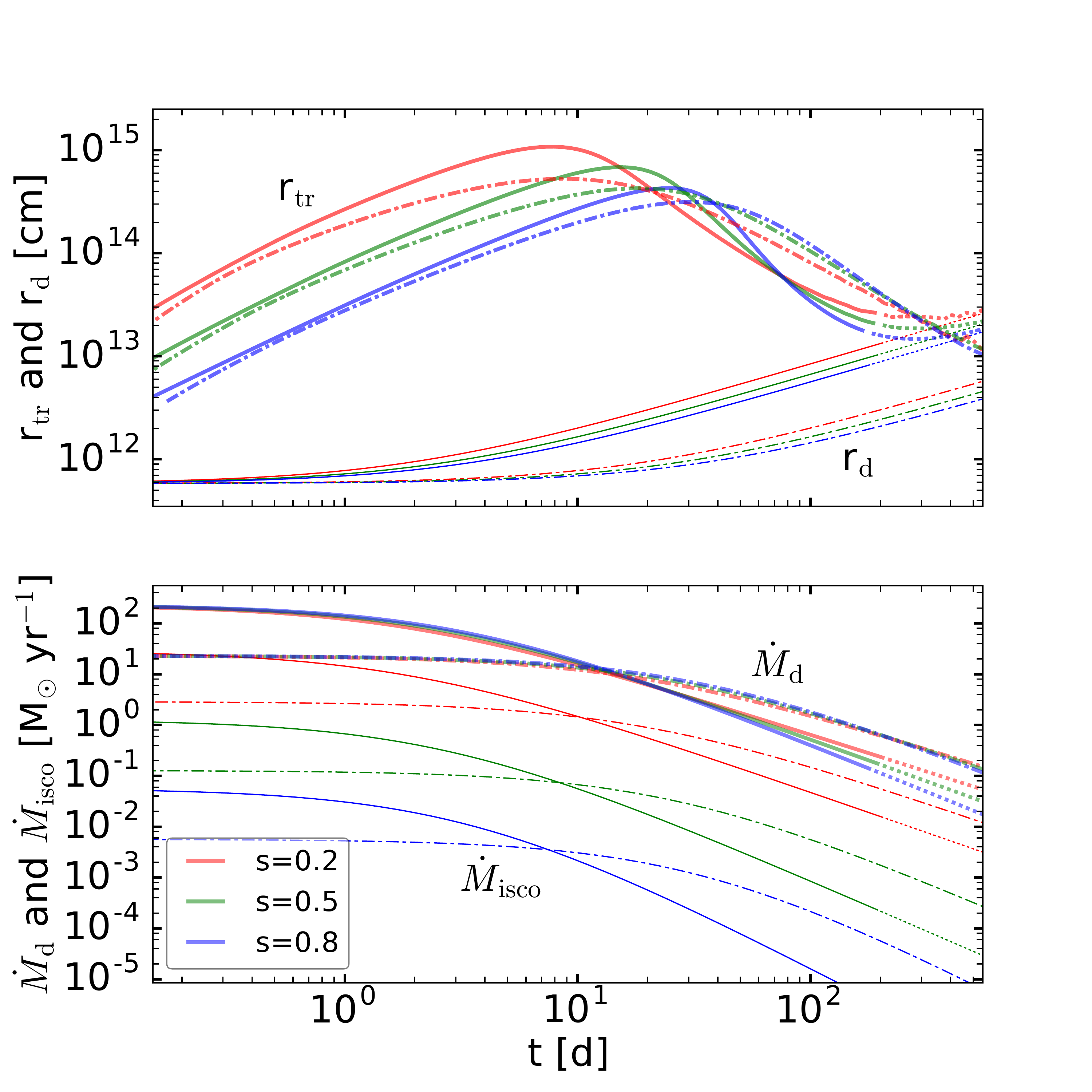}
\caption{The same as Figure \ref{fig:disk_evolution_0p5} but with $M_*=5\msun$. The major difference is that the disk accretion rate stays high and the disk stays thick for a longer time than than in the $M_*=0.5\msun$ case.
}\label{fig:disk_evolution_5}
\end{figure}

The electromagnetic signals of MS--BH TDEs depend on the physics of super-Eddington accretion, and the relevant MHD processes coupled with radiative transport are not understood in detail \citep{2011ApJ...733..110B, 2011ApJ...736....2O, 2014MNRAS.441.3177M, 2016MNRAS.456.3929S, 2017arXiv170902845J}, see recent reviews by \citet{2013LRR....16....1A, 2014SSRv..183...21B}. To construct a simple model, we approximate the disk mass distribution as a ``ring" located at the peak radius of the surface density distribution $\rd$ and calculate the time evolution of the bulk properties of the disk. The disk mass $M_{\rm d}$ and angular momentum $J_{\rm d}$ evolve as \citep{2008MNRAS.388.1729K}
\begin{equation}
    \label{eq:acc_rate}
    \dot{M}_{\rm d}(\rd) = -M_{\rm d}/t_{\rm vis},\ \dot{J}_{\rm d} = -f_{\rm j} J_{\rm d}/t_{\rm vis},
\end{equation}
where the viscous time is taken to be $t_{\rm vis}(\rd) = 4\alpha^{-1}\Omega_{\rm K}^{-1}(\rd)$ (fixing the height ratio $H/\rd = 0.5$) and the parameter $f_{\rm j}$ is the ratio between the specific angular momentum of the disk wind and that of the disk at the wind launching point. The disk angular momentum at any given time is given by $J_{\rm d} = (GM\rd)^{1/2}M_{\rm d}$, so the disk radius evolves as 
\begin{equation}
    \label{eq:disk_radius}
    \dot{r}_{\rm d} = 2(1-f_{\rm j})\rd/t_{\rm vis}.
\end{equation}
The analytical solution to equations (\ref{eq:acc_rate}) and (\ref{eq:disk_radius}) is
\begin{equation}
\label{eq:disk_evolution}
\begin{split}
    r_{\rm d} & = r_{\rm d,0} \left(1 + {t\over t_0}\right)^{2\over 3},\\
    \dot{M}_{\rm d} & = -{M_{\rm d,0}\over 3(1-f_{\rm j})t_0} \left(1 + {t\over t_0}\right)^{-{4-3f_{\rm j}\over 3(1-f_{\rm j})}},
\end{split}
\end{equation}
where $t_0 \equiv [4/3(1-f_{\rm j})]\alpha^{-1}\Omega_{\rm k}^{-1}(r_{\rm d,0})$ is the characteristic evolution timescale and the initial conditions for the disk evolution are taken as (cf. Section \ref{sec:DiskFormation})
\begin{equation}
    r_{\rm d,0} = \rt,\ M_{\rm d,0} = M_*/2.
\end{equation}
The time evolution of the disk accretion rate and radius for a number of choices of $f_{\rm j}$ and $\alpha$ are shown in Figures \ref{fig:disk_evolution_0p5} and \ref{fig:disk_evolution_5}.

It is widely believed that a large fraction of the disk mass will be blown away in the form of a wind \citep{1973A&A....24..337S,1994ApJ...428L..13N, 1999MNRAS.303L...1B, 2004MNRAS.349...68B, 2009MNRAS.400.2070S}. We approximate the radius-dependent accretion rate as a power-law \citep{1999MNRAS.303L...1B}
\begin{equation}
  \label{eq:8}
  \dot{M}(r) = \dot{M}(\rd) (r/\rd)^s, \ \risco < r< \rd,
\end{equation}
where the power-law index $0\leq s\leq 1$, and the upper limit of $s$ is set by energetic requirement and the lower limit corresponds to no wind mass loss. Numerical simulations of adiabatic accretion flows show that the power-law index likely lies in the range $0.4$--$0.8$ \citep[][and references therein]{2012ApJ...761..129Y}. The radius of the inner-most stable circular orbit (ISCO) depends on the BH spin $\risco\in[1, 9]\rg$. If the wind launched at radius $r$ has specific angular momentum $\t{f}_{\rm j} (GMr)^{1/2}$ (where $\t{f}_{\rm j}$ is the lever arm), then the rate at which angular momentum carried away by the wind is $\dot{J}_{\rm d} = -[\t{f}_{\rm j}s/(s + 1/2)]J_{\rm d}/t_{\rm vis}$ \citep{2014ApJ...784...87S}, which means $f_{\rm j} = \t{f}_{\rm j}s/(s + 1/2)$. The angular momentum evolution only affects the late-time behavior of the electromagnetic emission at time $t\gg t_{\rm vis}(r_{\rm d,0})$. Since we are mainly concerned with the peak luminosity and the peak duration, we take the lever-arm factor $\t{f}_{\rm j}\sim1$ for simplicity and hence
\begin{equation}
    f_{\rm j} \sim s/(s + 1/2).
\end{equation}
Under the above prescription, the system evolves self-similarly at $t\gg t_{\rm vis}(r_{\rm d,0})$ and we obtain the well-known solution: $\rd\propto t^{2/3}$, $\dot{M}_{\rm d} \propto t^{-(2s+4)/3}$ and $\dot{M}_{\rm isco} \propto t^{-4(s+1)/3}$ (see equation \ref{eq:disk_evolution}). The prescription stays valid until the accretion rate $\dot{M}_{\rm d}$ drops below $\sim (\rd/\rg)L_{\rm Edd}/c^2$  \citep[or the sphericalization radius $\dot{M}_{\rm d}c^2 \rg/L_{\rm Edd}$ drops below $\rd$,][]{1979MNRAS.187..237B}, and then the disk is expected to undergo thermal instability and collapse into a thin one near $\rd$. The accretion rate may drop by many orders of magnitude at the transition to thin-disk phase \citep{2014ApJ...784...87S}. In this paper, we focus on the thick-disk phase where an optically thick wind can generate bright optical emission. The time evolution of disk properties in the thick-disk phase are shown in Figures \ref{fig:disk_evolution_0p5} and \ref{fig:disk_evolution_5}.

\subsection{Radiation from the super-Eddington wind}
\label{sec:Radiation}
Having the disk dynamics in hand, we calculate the radiation-hydrodynamic evolution of the wind. The local viscous heating rate per unit area is given by \citep{1974MNRAS.168..603L}
\begin{equation}
  \label{eq:10}
  Q(r) = {3GM\dot{M}(r)\over 4\pi r^3} \left[1 - \left(\risco\over
      r\right)^{1/2}\right]. 
\end{equation}
We assume that a fraction $\feps$ of this heating power escapes\footnote{Note that $\feps$ is not much less than unity. For instance, if the super-Eddington wind launched at each radius has asymptotic velocity of order the local escape speed, then the wind kinetic power has efficiency
$\sim [s\dot{M}(r)/r][GM/r]/[2\pi r Q(r)] =(2s/3)[1 - \sqrt{\risco/r}]^{-1}$. }, shared by the wind kinetic power and radiative power. Then the total luminosity of the disk is given by
\begin{equation}
  \label{eq:11}
  L_{\rm d} = \feps \int_{\risco}^{\rd} Q(r) 2\pi r\d r = {3\feps 
    \rg\over 2\risco} \dot{M}(\rd) c^2 g(x, s),
\end{equation}
where $x \equiv\rd/\risco$ and the function $g$ is given by
\begin{equation}
  \label{eq:12}
  g(x, s) = x^{-s} \left({1-x^{s-1} \over 1-s} - {1-x^{s-1.5}
      \over 1.5-s} \right).
\end{equation}
Except for extreme values of $s$ (0 or 1), most mass is loaded near $\rd$ (lowest escape speed) but most accretion power is released near radius $\risco$ (highest escape speed), so we expect internal collisions to occur and the total power $L_{\rm d}$ will be thermalized near radius $\rd$ under spherical symmetry \citep{2012MNRAS.420.2912B}. Thus, the bulk kinetic energy and the radiation-dominated thermal energy are roughly in equipartition at $r\sim \rd$.

\begin{figure}
  \centering
\includegraphics[width = 0.48\textwidth,
  height=0.57\textheight]{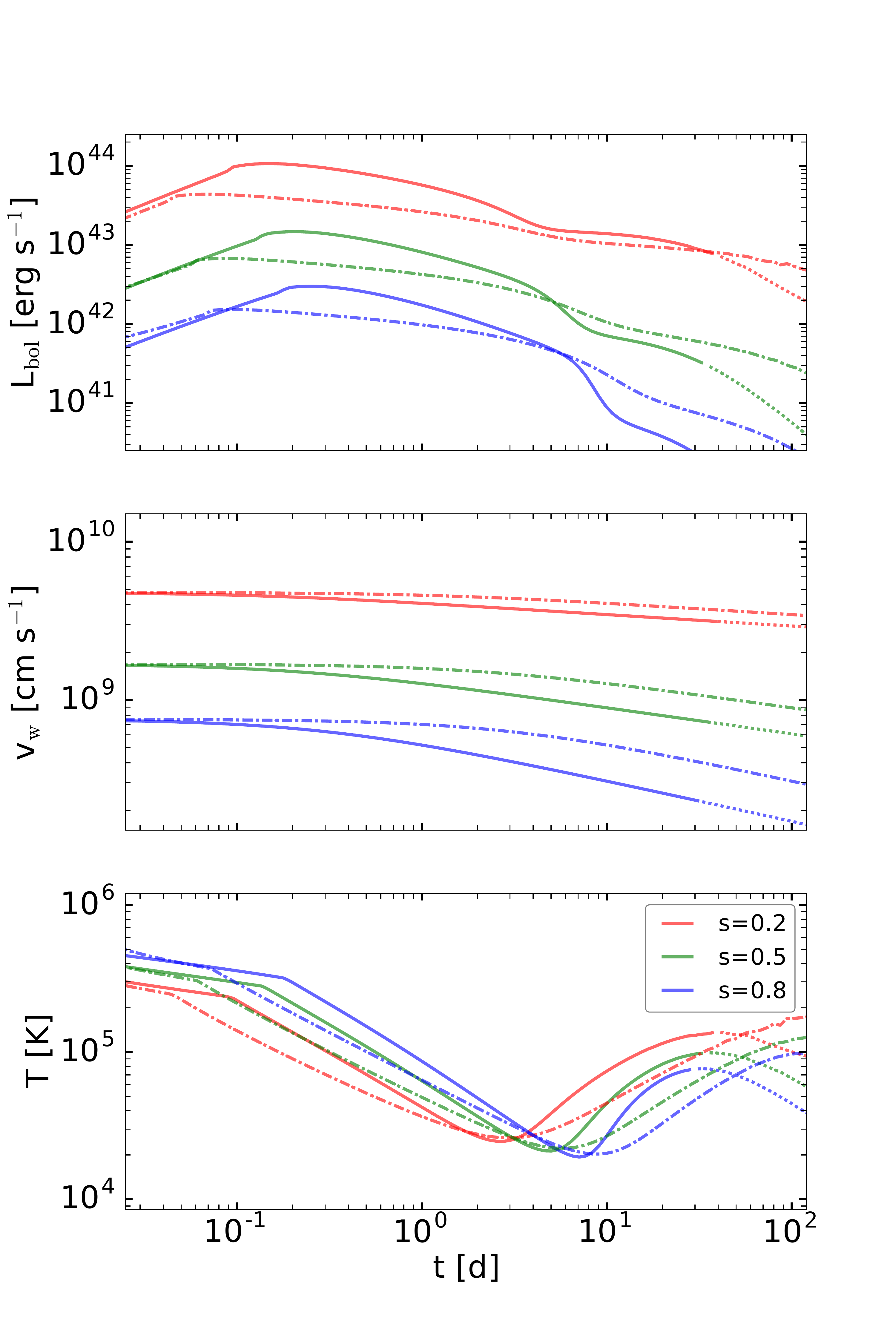}
\caption{Bolometric lightcurves (upper panel), asymptotic wind velocity (middle panel), and the temperature of the photons escaping from the trapping radius (lower panel), for three different choices of accretion rate power-law indexes $s=0.2$ (red), $0.5$ (green), and $0.8$ (blue).
Two cases of different viscosity parameters $\alpha$ are shown in solid curves ($\alpha=0.1$) and dash-dotted curves ($\alpha=0.01$).
We fix the efficiency of escaping power $\feps=0.5$ (equation \ref{eq:11}), ISCO radius $\risco=6\rg$, BH mass $M=20\msun$, and stellar mass $M_*=0.5\msun$. The solid (and dash-dotted) segment of each curve corresponds to when the accretion disk stays thick, and the dotted segment at late time indicates that the disk may collapse into a thin one (so our assumption of $H/r\sim 1$ becomes invalid).
}\label{fig:lc_0p5}
\end{figure}

As the wind expands in radius, nearly all the heat is converted back into bulk kinetic energy due to adiabatic expansion, so the asymptotic wind speed $\vw$ is given by $\dot{M}_{\rm w}\vw^2/2 = L_{\rm d}$, which means
\begin{equation}
  \label{eq:13}
  \vw/c = \left[{3\feps \rg \, g(x, s)
    \over  \risco (1-x^{-s})}\right]^{1/2}.
\end{equation}
Note that the asymptotic wind speed is independent of the accretion rate and is sub-relativistic in most situations (unless $s\approx 0$).

The radiation carried by the wind is advected until the photon-trapping radius $\rtr$ where photons can diffuse away over a dynamical time.
At a given time $t$, a wind shell at radius $r$ was launched at the retarded time $t_0(r)$ given by
\begin{equation}
    r = \rd(t_0) + \vw(t_0) (t - t_0),
\end{equation}
where $\rd(t_0)$ and $\vw(t_0)$ are the disk radius (inner boundary of the wind, equation \ref{eq:disk_evolution}) and the wind speed (equation \ref{eq:13}) at time $t_0$, respectively. Thus, the time evolution of the photon-trapping radius $\rtr(t)$ can be estimated by matching the diffusion time with the expansion time,
\begin{equation}
\label{eq:r_trap}
    {\tau(\rtr) \over c}{\rtr (r_{\rm out} - \rtr) \over r_{\rm out}} \simeq t - t_0(\rtr),
\end{equation}
where $t_0(r)$ is the retarded time of the wind shell currently at radius $r$, $r_{\rm out} = \vw(t_0=0) t$ is the outer boundary of the wind launched at $t_0=0$, and $\tau(r)$ is the Thomson scattering optical depth of the wind outside radius $r$,
\begin{equation}
    \tau(r) = \int_r^{r_{\rm out}} \kappa_{\rm s}\rho(r) \d r.
\end{equation}
The integrated wind mass loss rate is
\begin{equation}
  \label{eq:9}
  \dot{M}_{\rm w} = \dot{M}(\rd) - \dot{M}(\risco),
\end{equation}
and the wind density profile is taken to be
\begin{equation}
  \label{eq:15}
  \rho(r) = \dot{M}_{\rm w}(t_0)/[4\pi r^2\vw(t_0)],
\end{equation}
where we have taken into account wind propagation by using the retarded time $t_0(r)$ for each shell at radius $r$. At the relevant densities and temperatures, the Rosseland-mean opacity is dominated by Thomson scattering $\kappa_{\rm s} = 0.34\rm\,cm^2\,g^{-1}$ (for cosmic abundance). The time evolution of the photon-trapping radius is shown in Figures \ref{fig:disk_evolution_0p5} and \ref{fig:disk_evolution_5} for two TDEs with the same BH mass $M=20\msun$ but different stellar masses $M_*=0.5$ and $5\msun$, respectively. The qualitative result is that the photon-trapping radius increases with time initially (due to wind expansion), reaches a maximum that is many orders of magnitude larger than the disk radius (due to high optical depth), and then decreases rapidly as the wind mass-loss rate drops at late time.

Once we find the photon-trapping radius, the bolometric luminosity of the escaping photons is given by the rate at which radiation are advected across the trapping surface, i.e.
\begin{equation}
\label{eq:Lbol}
    L_{\rm bol}(t) = 4\pi \rtr^2 U(\rtr)\left[\vw(t_0) - {\d \rtr \over \d t}\right],
\end{equation}
where the radiation energy density is given by adiabatic expansion (valid at $r\leq \rtr$)
\begin{equation}
\label{eq:Urad}
    U(\rtr) = U(\rd) \left[\rho(\rtr)\over \rho(\rd)\right]^{4/3},
\end{equation}
and $U(\rd) = L_{\rm d}/(8\pi \rd^2 \vw)$ and $\rho(\rd) = \dot{M}_{\rm w}/(4\pi \rd^2 \vw)$ are evaluated using the disk luminosity $L_{\rm d}(t_0)$, disk radius $\rd(t_0)$, and asymptotic wind speed $\vw(t_0)$ at the retarded time for this trapping shell. The temperature (or average energy) of the escaping photons\footnote{The thermalization radius is defined where the effective optical depth $\tau_*(r_{\rm th})\simeq \sqrt{\tau_{\rm a}\tau}\simeq 1$ \citep{1979rpa..book.....R}, where $\tau_{\rm a}=\tau_{\rm s}\kappa_{\rm a}/\kappa_{\rm s}$ is the absorption optical depth. In the case when $\rtr<r_{\rm th}$, the color temperature is lower than that in equation (\ref{eq:T_escape}) by a factor of $(r_{\rm th}/\rtr)^{-3/4}$, but the bolometric luminosity in equation (\ref{eq:Lbol}) is unaffected. Modeling the (bound-free and free-free) absorption opacity $\kappa_{\rm a}$ requires detailed non-LTE radiative transfer calculations and is left for future works.} is given by
\begin{equation}
\label{eq:T_escape}
    T(\rtr) = \left[U(\rtr)c/4\sigma_{\rm SB}\right]^{1/4},
\end{equation}
where $\sigma_{\rm SB}$ is the Stefan-Boltzmann constant. 

\begin{figure}
  \centering
\includegraphics[width = 0.48\textwidth,
  height=0.57\textheight]{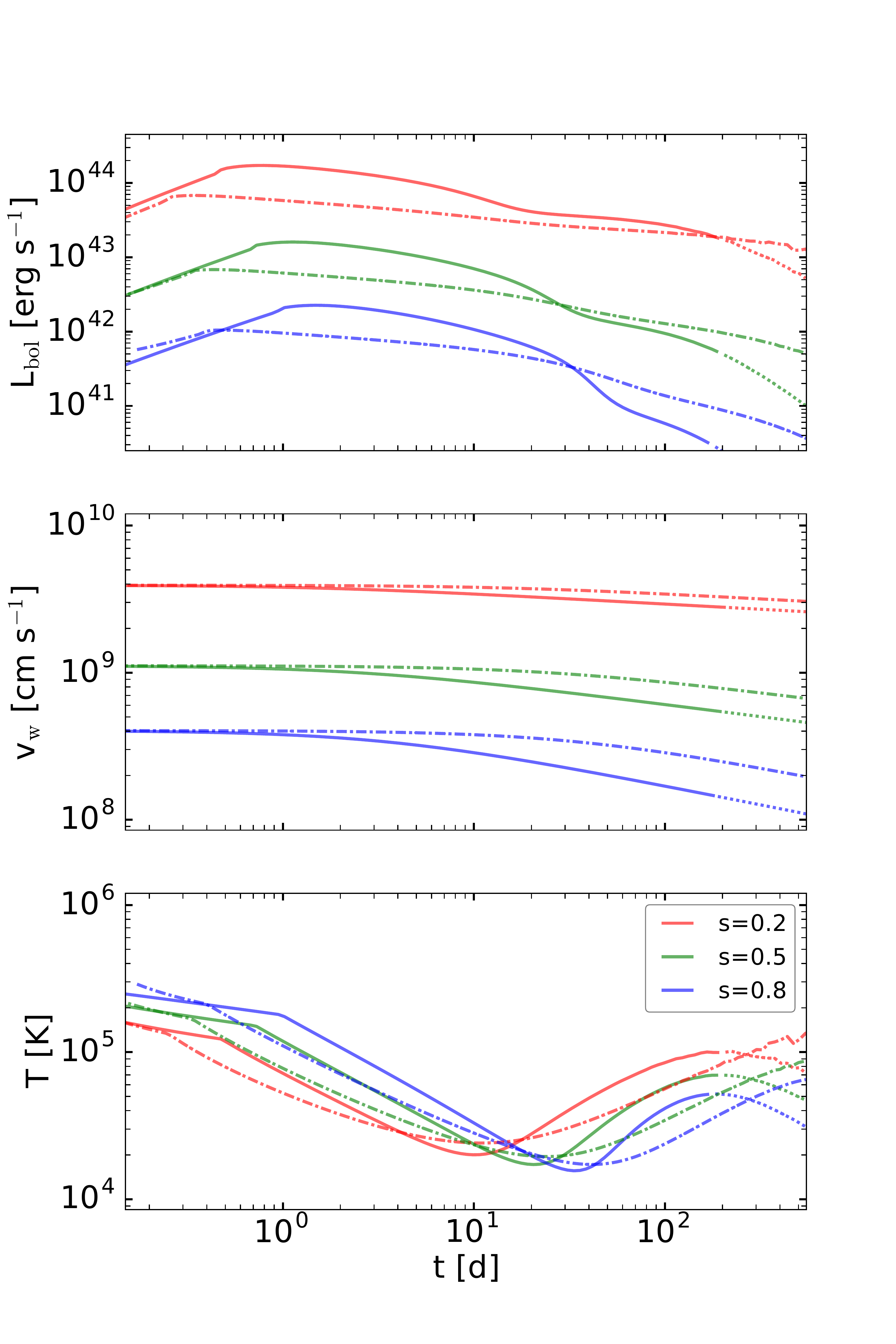}
\caption{The same as Figure \ref{fig:lc_0p5} but with $M_*=5\msun$. The main difference is that the luminosity is slightly higher, the photon temperature is slightly lower, and the emission lasts for a longer time than in the $M_*=0.5\msun$ case.
}\label{fig:lc_5}
\end{figure}

Figures \ref{fig:lc_0p5} and \ref{fig:lc_5} show the bolometric lightcurve, asymptotic wind velocity, and the temperature of the photons escaping from the trapping surface. We find that BH-MS TDEs generate bright optical transients on timescales of a few to 100 days with peak luminosities from $10^{41}$ to $10^{43}\rm\,erg\,s^{-1}$. The photospheric velocity may range from 0.01$c$ to 0.1$c$ and decreases with time. The largest uncertainty in our model is the mass flux power-law index $s$, and the luminosity decreases towards larger $s$ (as a result of stronger mass loss and hence less accretion power). We note that our simple model for the disk evolution does not include the rising segment of the wind power at very early time, which depends on the details of the tidal disruption and disk formation on a timescale of a few hours. Instead, we add a short segment\footnote{A steeper power-law generally leads to a more rapidly rising lightcurve, but the latter is much shallower in that $\d \mr{log} L/\d \mr{log} t < \d\mr{log} \dot{M}(\rd)/\d \mr{log} t$, because of the lag between wind launching and photon escaping. The lag time is longer for larger mass flux index $s$ due to a denser wind.} of $\dot{M}(\rd)\propto t^5$ in the first $10^3\,$sec (or $10^4\,$sec) for the case of $M_*=0.5\msun$ (or $M_*=5\msun$), which roughly captures the rapid onset of the gas fallback as shown in Figure 1 of \citet{Perets2016}. Input from three-dimensional SPH calculations is needed to discuss the intricacies of disk formation and evolution in more detail in the context of the calculations presented here.

\subsection{Jet formation and gamma-ray burst}
\label{sec:GRB}

Given the high peak accretion rate (equation \ref{eq:7}), if the accretion efficiency can reach that of typical active galactic nuclei $\sim0.1$ \citep[e.g.][]{1982MNRAS.200..115S}, then the peak luminosity of the accretion system is of order $10^{48}\rm\,erg\,s^{-1}$ with duration $\sim 10^4$--$10^5\,$sec. Thus, it has been proposed \citep{Perets2016} that TDEs by stellar-mass BHs may be responsible for the population of ultra-long duration gamma-ray bursts \citep[ulGRBs,][]{2013ApJ...766...30G, 2014ApJ...781...13L}. Numerical simulations of super-Eddington accretion flows in the magnetic-arrested disk (MAD, with dynamically important magnetic flux and rapid BH spin) regime show that high accretion efficiency $> 0.1$ can be achieved \citep{2011MNRAS.418L..79T, 2015MNRAS.454L...6M}, and these models have been applied to ultra-luminous X-ray sources \citep{2017MNRAS.469.2997N} and jetted TDEs by super-massive BHs \citep{2014MNRAS.437.2744T, Dai2018, 2019MNRAS.483..565C}.

\citet{2014ApJ...781...13L} estimated the event rate of ulGRBs to be a factor of a few lower than that of classical GRBs, of order $1\rm\,Gpc^{-3}\,yr^{-1}$ at redshift $z\sim 1$, after correcting for the selection bias that faint, long-lived events tend to fall below the trigger threshold of \textit{Swift} BAT. However, ulGRBs are most likely strongly beamed, so the true rate may be much (a factor of 10--100) higher than the BH-MS rate predicted by our GC simulations (see Figure \ref{fig:rate}). Another challenge to the BH-MS TDE scenario is the detection of a bright supernova (SN2011kl) associated with the ultra-long GRB 111209A \citep{2015Natur.523..189G}. SN2011kl had a peak time of 14 rest-frame days and peak bolometric luminosity of $\sim3 \times 10^{43} \rm\,erg\,s^{-1}$, and its overall spectral and lightcurve shapes were similar to that of GRB-associated broad-line type-Ic supernovae (with no evidence of H or He). This is inconsistent with the expected H-rich gas composition of BH-MS TDEs, although the super-Eddington disk wind may generate sufficiently bright optical emission (see the $s=0.2$ case in Figure \ref{fig:lc_0p5}).

Therefore, we conclude that it is unlikely that BH--MS TDEs are responsible for the majority of ulGRBs. Nevertheless, when appropriate conditions (e.g. BH spin and magnetic flux) are met, the launching of relativistic jets in BH--MS TDEs is still possible. In the following section, we compute the comoving and cumulative rates of BH--MS TDEs as a function of redshift, showing that, in principle if indeed a relativistic jet is launched, these events can potentially be detectable at very high redshifts.

An alternative possibility mentioned briefly in \citet{Perets2016} is that, on longer timescales, debris from the TDE may slowly fall back, potentially forming a long-lived accretion disk that may produce an X-ray source very similar to an X-ray binary (with the notable difference that the accretion occurs through the disk alone, not fed by a stellar companion). We refer the reader to previous work by \citet{Hills1976} and \citet{Krolik1984} for a more detailed discussion of this scenario.

\section{Event Rates}
\label{sec:rates}

\begin{figure}
\begin{center}
\includegraphics[width=0.45\textwidth]{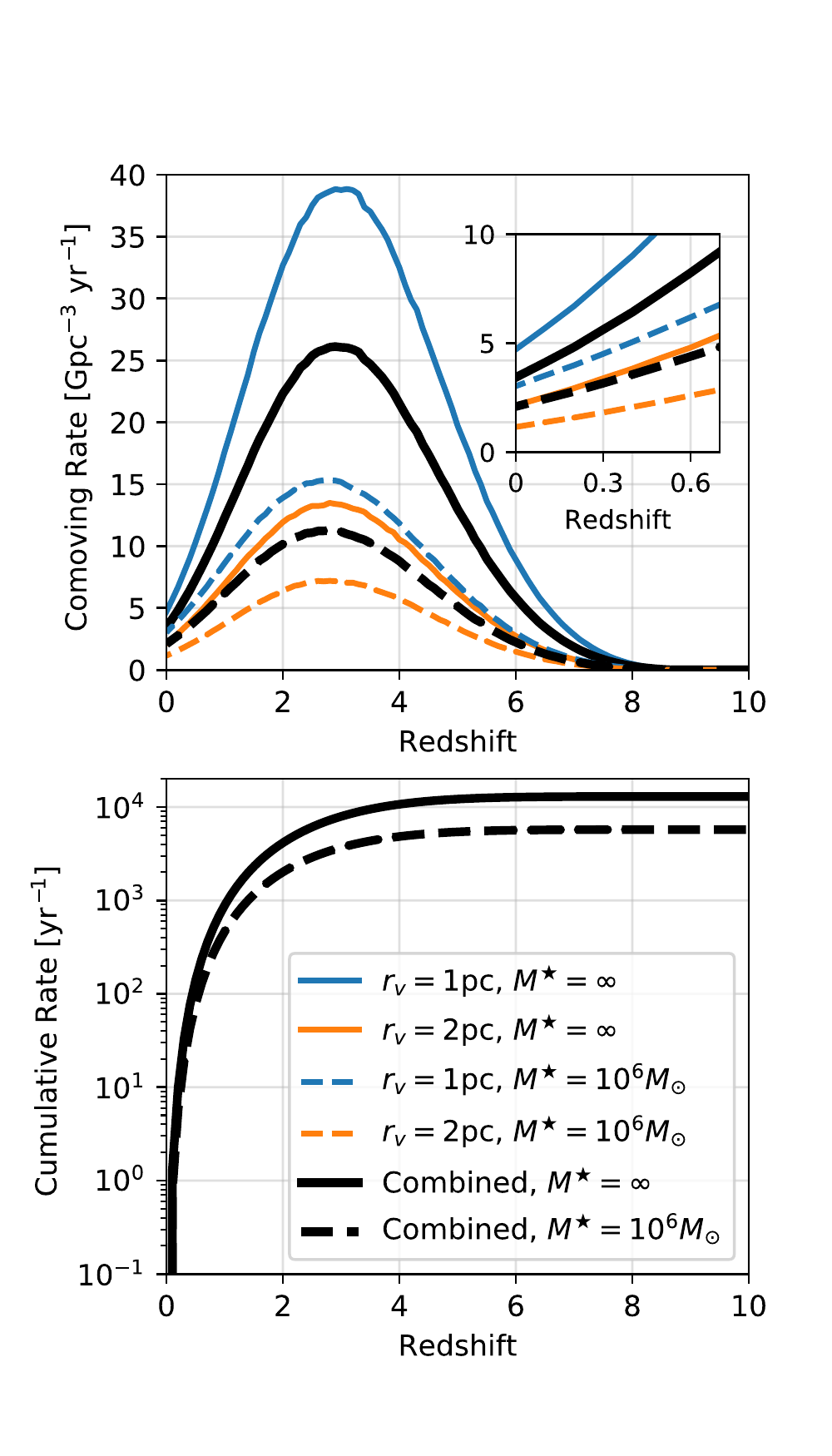}
\caption{\label{fig:rate} Comoving rate (top panel) and cumulative rate (bottom panel) versus redshift for all BH--MS TDEs, as calculated using the method described in Section \ref{sec:rates}. In blue (green) we show the rate for clusters of $r_v=1\,(2)\,$pc, and in black we show the combined rate. Solid lines show rates assuming our fiducial cluster initial mass function (a pure $\propto M^{-2}$ power law with no truncation) and dashed lines show }
\end{center}
\end{figure}

To calculate the event rate of BH--MS TDEs, we use an integral equation similar to that used in \citet{RodriguezLoeb2018}. The BH--MS TDE rate as a function of cosmic time $t$ is given by

\begin{multline}
\label{eq:rate_calculation}
\mathcal{R}(t) = \iiint \frac{\dot{M}_{\rm{GC}}}{d \log M_{\rm{Halo}}} \bigg\rvert_{z(\tau)}\frac{1}{\langle M_{\rm{GC}} \rangle} P(M_{\rm{GC}}) \\ \times R(r_v,M_{\rm{GC}},\tau-t)f_{\rm no IMBH}(r_v,M_{\rm GC}) dM_{\rm{Halo}} dM_{\rm{GC}}d\tau
\end{multline}

  \noindent where  $\frac{ \dot{M}_{\rm GC}} {d\log_{10}M_{\rm    Halo}}$ is the 
	      comoving rate of star formation in GCs (in units of $M_{\odot} {\rm yr}^{-1} {\rm Mpc}^{-3} $) per 
	       halo mass $M_{\rm Halo}$ at redshift $z(\tau)$ corresponding to a cosmic time $\tau$, from \citet{El-Badry2018}.  $R(r_{v},M_{\rm GC},t)$ is the rate of BH--MS TDEs at time for for a cluster with an initial mass $M_{\rm GC}$ and virial radius $r_{v}$.  As was done in \citet{RodriguezLoeb2018}, we assume the individual cluster rate can be described as

	         \begin{equation}
	  R(M_{\rm GC},t) \equiv  (A M_{\rm GC}^2 + B M_{\rm GC} + 
	  C) \times t^{-(\gamma + \gamma_M \log_{10}M_{\rm GC})}
	  \label{eqn:gcrate}
  \end{equation}
  
  \noindent where we fit the 5 parameters, $\theta = (A,B,C,\gamma,\gamma_M)$, separately for clusters with $r_v = 1\rm{pc}$ and $r_v = 2\rm{pc}$.  We truncate the rate to zero below 100 Myr, which we find reproduces both the rate and total number of BH--MS TDEs from our CMC models.

       $P(M_{\rm GC})$ is the cluster initial mass function (CIMF), which 
	      we assume to be a power-law between $10^5M_{\odot}$ and $10^7M_{\odot}$, with a possible exponential truncation
	      
	        \begin{equation}
	  \phi(M_{\rm GC})dM_{\rm GC} \propto M_{\rm GC}^{-2} \exp( - M_{\rm GC} / 
	  M_{\rm GC}^{\bigstar})dM_{\rm GC}~.
	  \label{eqn:schechter}
  \end{equation}
  
  \noindent We consider exponential truncations of $M_{\rm GC}^{\bigstar} = \infty$ (corresponding to a purely $\propto M^{-2}$ power law) and $M_{\rm GC}^{\bigstar} = 10^6M_{\odot}$ (as suggested by observations of young massive star clusters in the local universe, e.g. \cite{PortegiesZwart2010}).  $\left< M_{\rm GC}\right>$ is the mean initial mass of GCs given an assumed CIMF (this is used to convert the mass formed in GCs into a number of GCs).   $\left< M_{\rm GC}\right>$ is the mean initial mass of GCs given the assumed CIMF.
  
  We have introduced a new parameter, $f_{\rm no IMBH}(r_v,M_{\rm GC})$, corresponding to the fraction of clusters that do not form an IMBH by runaway collisions of BHs.  It has recently been shown \cite{Antonini2018} that for clusters with central escape speeds $\gtrsim 300 \rm{km}/\rm{s}$ the probability of forming an IMBH through repeated collisions of BHs goes to 1.  Since the presence of an IMBH can suppress the standard three- and four-body encounters that facilitate collisions between BHs and MS stars, we assume that any cluster which forms an IMBH does not contribute to the rate calculated. This is incorporated into equation~\eqref{eq:rate_calculation} by assuming that a certain fraction of clusters with a given escape speed $v_{\rm esc}$ contribute zero mergers.  We find that an equation of the form
  
  \begin{equation}
    f_{\rm no IMBH}(r_v,M_{\rm GC}) =  \left[ 1 + \exp(k  (v_{\rm esc} - v_{\rm esc}^{\bigstar})) \right]^{-1}~,
    \label{eqn:imbh}
  \end{equation}
  
  \noindent with $k = 0.05$ and $v_{\rm esc}^{\bigstar} = 295 \rm{km}/\rm{s}$, provides a good fit to Figure 6 of \cite{Antonini2018}.  We define $v_{\rm esc}$ as the central escape speed from a Plummer model of mass $M_{\rm GC}$ and virial radius $r_v$ \citep[e.g.,][]{HeggieHut2003},
  
  \begin{equation}
  v_{\rm esc} = \sqrt{\frac{32GM_{\rm GC}} {3\pi r_v}}~.
  \end{equation}
  
  \noindent In practice, equation~\eqref{eqn:imbh} only truncates the contribution to the rate from the most massive and compact clusters.  We find that the local comoving merger rate is decreased by only $\sim 10\%$ in the local universe.  For comparison, this same correction decreases the BBH merger rate from \cite{RodriguezLoeb2018} from $\sim 14 \,{\rm Gpc}^{-3} {\rm yr}^{-1}$ to $\sim 12\,{\rm Gpc}^{-3} {\rm yr}^{-1}$ in the local universe.  The IMBH correction does not alter the rate when the CIMF is truncated by $M^{\bigstar}_{\rm GC} = 10^6M_{\odot}$. We discuss the topic of IMBHs further in Section \ref{sec:future}.

Figure \ref{fig:rate} shows the rate as a function of redshift using equation \ref{eq:rate_calculation}.  The top panel shows the comoving rate and the bottom panel shows the cumulative rate. The solid and dashed black lines show the BH--MS TDE rate for both the fiducial CIMF (solid) and assuming an exponential truncation at $M^{\star}=10^6\,M_{\odot}$ (dashed). In the top panel, we also show in blue (green) the rate if only those models with $r_v=1\,$pc ($2\,$pc) are considered in the calculation. As anticipated, the predicted rate is higher for the models with $r_v=1\,$pc, simply because these models have relatively higher stellar densities, leading to more TDEs.

As discussed in Section~\ref{sec:method}, if we consider our upper limit for the TDE cross section, the rate shown in Figure~\ref{fig:rate} by a factor of roughly 3, based on our typical values of BH and MS masses. Therefore, we can extrapolate an upper limit for the BH--MS TDE rate by simply multiplying the rates of Figure \ref{fig:rate} by this factor. In the local universe ($z<0.1$), we predict a comoving BH--MS TDE rate of $4\,\rm{Gpc}^{-3}\rm{yr}^{-1}$ and an extrapolated upper limit rate of $12\,\rm{Gpc}^{-3}\rm{yr}^{-1}$. If we consider a CIMF truncated at $10^6\,M_{\odot}$, these rates decrease to $2.5\,\rm{Gpc}^{-3}\rm{yr}^{-1}$ and $7.5\,\rm{Gpc}^{-3}\rm{yr}^{-1}$, respectively. For comparison, \citet{RodriguezLoeb2018} predicted a typical BBH merger rate of $12\,\rm{Gpc}^{-3}\rm{yr}^{-1}$ in the local universe, roughly comparable to our extrapolated upper limit TDE rate if we assume no truncation for the CIMF.

Assuming the density of Milky Way equivalent galaxies (MWEGs) in the local universe is one per $(4.4\,\rm{Mpc})^3$ \citep{Abadie2010}, we predict a BH--MS TDE rate of $3.4\times 10^{-7}\,\rm{MWEG}^{-1}\rm{yr}^{-1}$ with an upper limit of roughly $10^{-6}\,\rm{MWEG}^{-1}\rm{yr}^{-1}$ in the local universe. Note that the latter rate is in approximate agreement with the analytic predictions made by \citet{Perets2016}.

\section{BH--giant close encounters}
\label{sec:giants}

\begin{figure}
\begin{center}
\includegraphics[width=0.5\textwidth]{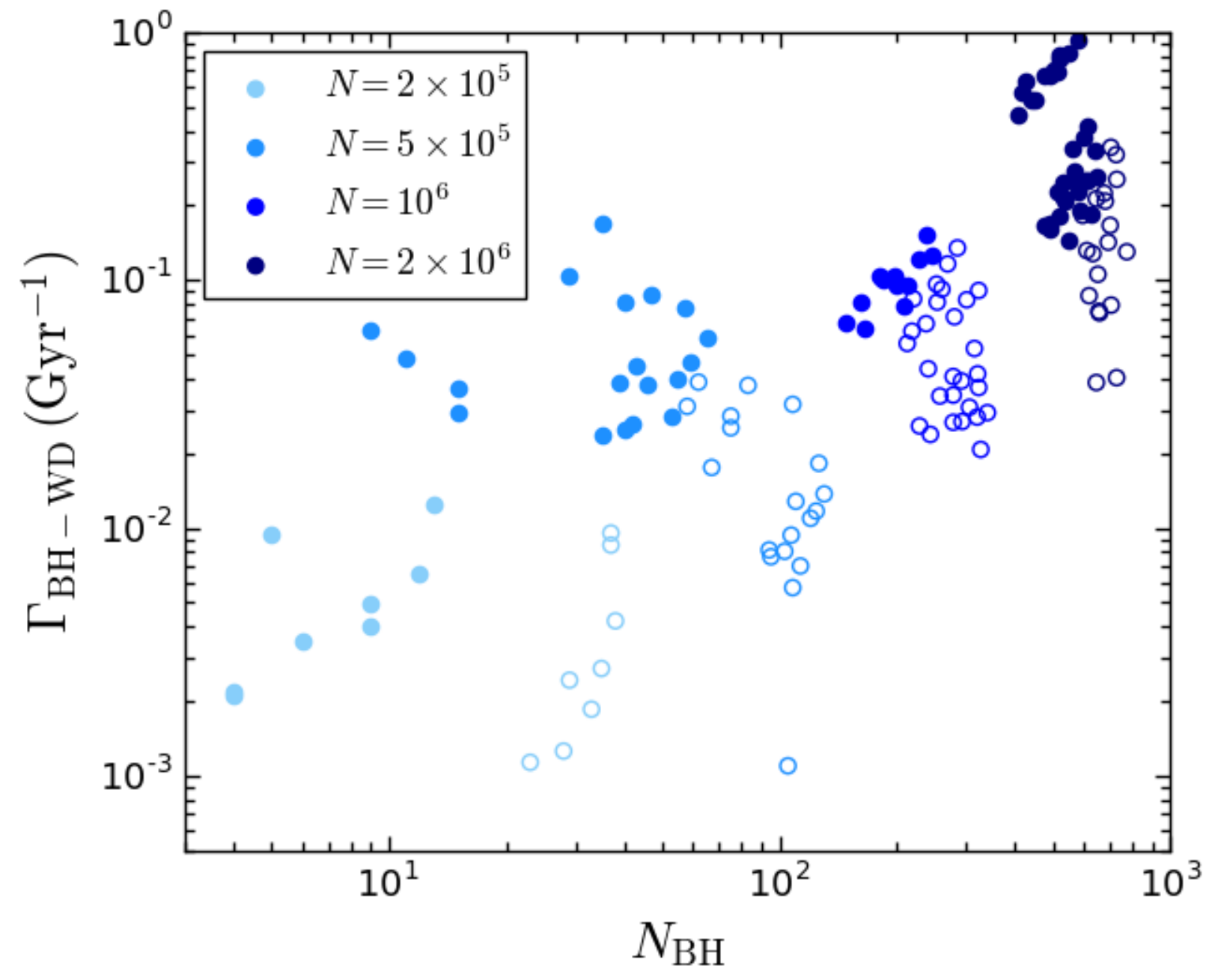}
\caption{\label{fig:giants} On the y-axis we show the formation rate of BH--WD binaries through BH--giant close encounters (as computed from equation \ref{eq:rate_BHWD}) at various late-time snapshots for models 1-24 and on the x-axis we show the total number of BHs retained in each cluster at these snapshots. Different colors denote different size clusters, as denoted in the legend. Filled circles show cluster models with $r_v=1\,$pc and open circles show models with $r_v=2\,$pc.}
\end{center}
\end{figure}

In addition to the BH--MS TDEs which are the main focus of this study, we also report all close encounters of BH--giant pairs identified in our models. The orange points in the top panel of Figure \ref{fig:scatter} show these events. In total, we find 246 giant TDEs in models 1-24. Although the cross section is higher for giants compared to MS stars (simply due to their relatively large stellar radii) the TDE rate is limited by the relatively small number of giants, which is a consequence of their shorter lifetimes.

Detailed consideration of the outcome of BH--giant close encounters is outside the scope of this study. We simply cite here the predicted outcomes from earlier work, and reserve a more detailed study of these events for a future analysis. In particular, \citet{Ivanova2010} showed that close encounters of giants and stellar-mass BHs may serve as a formation channel for BH--WD binaries, an important result given that observations suggest that a fraction of the stellar-mass BH candidates identified in GCs to date may indeed have WD companions \citep{Strader2012,Bahramian2017}. Estimates from \citet{Ivanova2010} of the hydrodynamics suggest that BH--giant collisions will result in a BH--WD binary with orbital separation $a \gtrsim 1.3\,R_{\rm{G}}$, where $R_G$ is the giant's radius. For giants of mass $\approx1\,M_{\odot}$ and radii $\approx 3\,R_{\odot}$ this corresponds to $a \gtrsim 10^{-2}\,$AU. Of course, the details of these BH--giant close encounters are sensitive to the highly uncertain common envelope physics and more detailed (hydrodynamic) calculations are necessary to explore the outcomes of these events in detail.

For typical clusters (number densities of $\sim 10^5\,\rm{pc}^{-3}$ and velocity dispersions of $\sim10\rm{km\,s}^{-1}$), \citet{Ivanova2010} predicted BH--WD binaries will form through BH--giant close encounters at a rate of $\sim 1.5 \times 10^{-2}$ per BH per Gyr, using a calculation of the form:

\begin{multline}
\label{eq:rate_BHWD}
\Gamma_{\rm{BH-WD}} \approx 0.1 f_{\rm{G}} f_p \Bigg(\frac{M_{\rm{BH}}}{15\,M_{\odot}} \Bigg) \Bigg(\frac{n}{10^5\,\rm{pc}^{-3}} \Bigg) \Bigg( \frac{R_{\rm{G}}}{R_{\odot}} \Bigg) \\ \times \Bigg(\frac{10\,\rm{km\,s}^{-1}}{\sigma_v} \Bigg)\times N_{\rm{BH}}\,\rm{Gyr}^{-1}
\end{multline}
\citep[see, e.g.,][]{Ivanova2005,Ivanova2010} where $f_{\rm{G}}$ is the fraction of giants in the stellar population within the GC core, $n$ is the number density, $M_{\rm{BH}}$ is the typical BH mass, $R_{\rm{G}}$ is the typical stellar radius of giants, and $\sigma_v$ is the velocity dispersion. $f_p=r_p/R_{\rm{RG}}$ describes how close a typical BH--giant encounter must be to result in a disruption. In our models, we take $f_p=1$, but as discussed in \citet{Ivanova2010}, $f_p$ as high as $\approx 5$ may also be appropriate.

In Figure \ref{fig:giants}, we show rate of formation of BH--WDs through BH--giant close encounters computed for various late time cluster snapshots ($t>8\,$Gyr) in models 1-24 plotted against the total number of BHs retained in the cluster at each respective time. We compute the rate, $\Gamma_{\rm{BH-WD}}$ using equation \ref{eq:rate_BHWD} with $M_{\rm{BH}}$, $R_{\rm{G}}$, $n$, $\sigma_v$, and $f_{\rm{G}}$ computed uniquely for each cluster snapshot. As in Figure \ref{fig:models}, filled circles denote models with $r_v=1\,$pc and open circles denote models with $r_v=2\,$pc. We limit ourselves to late times here to reflect the typical ages of present-day GCs. BH--WD binaries formed through close encounters of BHs and giants at earlier times may be unlikely to survive to the present as a result of the frequent dynamical encounters which may break apart the binaries,\footnote{The typical timescale for a BH--WD binary to undergo a strong encounter that could potentially disrupt the binary can be approximated as $t \sim 3 \mathrm{Gyr} \,(n/10^5\,\mathrm{pc}^{-3})^{-1}(a/0.1\,\mathrm{AU})^{-1} (M/30\,M_{\odot})^{-1}(\sigma_v/10\,\mathrm{km/s})$. Thus we expect a typical BH--WD binary to survive for no longer than a few Gyr in a typical cluster. See, e.g., \citet{Kremer2018a} for further discussion of the rate at which BH--non-BH binaries are expected to be broken through dynamical encounters.} thus to remain conservative we include only those binaries that would have formed relatively recently in their host cluster.

Figure \ref{fig:giants} shows similar trends to those shown in Figure \ref{fig:models} for the BH--MS TDEs. In particular, the formation rate of BH--WDs varies directly with $N$ (as seen by comparing the different colors) and inversely with cluster size ($r_v$; as seen by comparing the open versus filled circles), as expected from equation \ref{eq:rate_BHWD}. The dependence of the rate on the total number of retained BHs is more complex. The overall trend is a direct relation between $\Gamma_{\rm{BH-WD}}$ and $N_{\rm{BH}}$, as expected because more BHs are formed in more massive clusters. However, for models of fixed mass (i.e., points of a single color in Figure \ref{fig:giants}), it can be seen that $\Gamma_{\rm{BH-WD}}$ varies \textit{inversely} with $N_{\rm{BH}}$.  This results from the way BH populations shape the dynamical evolution of their host cluster. As discussed in \citet{Kremer2019a}, BH populations provide an internal ``heating'' source for their host cluster. As a result, clusters with many BHs are relatively diffuse while clusters with few BHs may undergo core-collapse leading to relatively high central densities. Because $\Gamma_{\rm{BH-WD}}$ scales with density, this means clusters with smaller BH populations will actually lead to \textit{more} BH--giant encounters. Furthermore, as the figure shows, this process is coupled with the initial $r_v$: clusters with $r_v=1\,$pc (filled circles) retain fewer BHs at late times compared to their $r_v=2\,$pc counterparts (open circles). This is also consistent with \citet{Kremer2019a}, who showed that the initial cluster size is the key parameter for determining the number of BHs retained in a cluster at present.

As Figure \ref{fig:giants} shows, close encounters of BHs and giants may lead to the formation of up to approximately one BH--WD binary per Gyr per cluster, possibly sufficient to explain the handful of accreting BH--WD binary candidates observed in local clusters. In addition to the BH--giant encounter channel considered here, BH--WD binaries compact enough to start mass transfer may also form through series of binary-mediated exchange encounters, as discussed in \citet{Kremer2018a}.

Finally, we note that, in addition to being potentially observed as mass-transferring low-mass X-ray binaries, BH--WD binaries may also be observed as gravitational wave (GW) sources by low-frequency GW detectors such as LISA \citep[e.g.,][]{Kremer2018c}. Future electromagnetic and GW observations of these sources will continue to provide better constraints on the formation of these types of binaries.

\section{Conclusions and Discussion}
\label{sec:discussion}

\subsection{Summary}
We have explored the disruption of stars by stellar-mass BHs in GCs through close encounters. We summarize our main findings below.

\begin{enumerate}
    \item Using our Monte Carlo code \texttt{CMC} to model the evolution of GCs, we show that stellar-mass BHs disrupt MS stars, giants, and WDs throughout the lifetime of the cluster. These TDEs can occur through both single--single and binary-mediated encounters (binary--single and binary--binary).
    \item The number of TDEs per cluster is determined primarily by the cluster's initial size (parameterized in our models by the initial virial radius) and initial mass. For our most massive and compact models ($N=2\times10^6$ and $r_v=1\,$pc), we get up to 200 BH--MS TDEs over the cluster lifetime. For our lower mass cluster models ($N = 2\times 10^5$), the total number of TDEs can be as low as zero to a few.
    \item By incorporating a realistic cosmological model for GC formation, we derive a rate of BH--MS TDEs of approximately $4-12\,\rm{Gpc}^{-3}\,\rm{yr}^{-1}$ in the local universe and  a cosmological rate that peaks at roughly $25-75\,\rm{Gpc}^{-3}\,\rm{yr}^{-1}$ at a redshift of about~3. %Notably, this is comparable to the typical binary BH merger rate from clusters of $14\,\rm{Gpc}^{-3}\,\rm{yr}^{-1}$predicted by \citet{RodriguezLoeb2018}.
    \item We show that the wind mass loss associated with these BH--MS TDEs produces optical transients of luminosity $\sim10^{41}$ to $10^{44}\rm\,erg\,s^{-1}$ on timescales of about a day to a month. In Section~\ref{sec:electromagnetic} we show lightcurve predictions expected from these events.
    \item BH--giant close encounters occur at rates of up to  $\sim1$ per Gyr per cluster. These events may serve as a dynamical formation channel for BH--WD binaries in GCs, which may be observed as X-ray or gravitational wave sources.
\end{enumerate}

\subsection{Detectability}
The Zwicky Transient Facility (ZTF) reaches $g$ = 20.8 mag (5$\sigma$) during a single exposure of $30\,$s, surveying $3\pi$ of the sky. Assuming a blackbody temperature of $2\times10^4\,$K (see Figures \ref{fig:lc_0p5} and \ref{fig:lc_5}), the detection horizon in luminosity distance is roughly $D_{\rm L,max}\simeq 150L_{\rm bol,42}\rm\,Mpc$, where  $L_{\rm bol} = 10^{42}L_{\rm bol,42}^{1/2}\rm\,erg\,s^{-1}$ is the bolometric luminosity near the optical peak. The all-sky rate is about $10\rm\,yr^{-1}$ in the optimistic case where $s=0.2$ and peak luminosity $L_{\rm bol}\sim 10^{44}\rm\,erg\,s^{-1}$. In the pessimistic case where $s=0.8$ and peak luminosity $L_{\rm bol}\sim 10^{42}\rm\,erg\,s^{-1}$, the all-sky rate is about $10^{-2}\rm\,yr^{-1}$. 

\subsection{Possible effects on the BH population}
\label{sec:effects}

When a BH disrupts a star, the disruption may have various effects upon the BH itself, which in turn may alter the overall BH dynamics in the cluster. For instance, if a significant fraction of the mass of the disrupted star were accreted by the BH following the TDE, the BH may be spun up through accretion. Merging BBHs that are highly spinning can get gravitational wave recoil kicks as high as $5000\,\rm{km\, s^{-1}}$ \citep[e.g.,][]{Campanelli2007,Lousto2012}, significantly larger than the escape speed of a typical GC or even a galactic nucleus. Thus, if a significant number of BHs in a given cluster attain high spins through TDEs of MS stars, all BBH merger products will be ejected from the cluster promptly after merger. This has important implications for the production of second-generation mergers in clusters \citep[see, e.g.,][]{Rodriguez2018a}.

Additionally, as discussed in Section \ref{sec:electromagnetic}, in a typical BH--MS TDE, some fraction of stellar material will become unbound from the system. This unbound mass is ejected in an asymmetric manner, so the BH receives an impulsive kick in response. The unbound debris has positive specific energies in the range $\eps \in (0, \eps_{\rm{max}})$, where $\eps_{\rm{max}}=v_{\rm{max}}^2/2$ is given by

\begin{equation}
\eps_{\rm{max}} \simeq \frac{G M_{\rm{BH}}}{\rt^{\,\,2}} R_{\star} = \Big(\frac{M_{\rm{BH}}}{M_{\star}} \Big)^{1/3} \frac{G M_{\star}}{R_{\star}}
\end{equation}
where we have used $\rt=R_{\star} (M_{\rm{BH}}/M_{\star})^{1/3}$. The escape velocity of the star is given by $v_{\rm{esc}}=\sqrt{2GM_{\star}/R_{\star}}$. The maximum speed of the unbound debris is then $v_{\rm{max}} = (M_{\rm{BH}}/M_{\star})^{1/6} v_{\rm{esc}}$. The total linear momentum, $P$, carried away by the unbound debris depends on the mass distribution over specific energy, $\rm{d}M/ \rm{d} \eps$, but a rough estimate is that $P \simeq (M_{\star}/3)v_{\rm{max}}$ (this expression is exact for a flat distribution of $\rm{d}M/ \rm{d} \eps$). Then the BH receives a kick in the $-P$ direction of velocity

\begin{equation}
v_{\rm{kick}} \simeq \frac{M_{\star} v_{\rm{max}}}{3M_{\rm{BH}}} \simeq \Big( \frac{M_{\star}}{M_{\rm{BH}}}\Big)^{5/6} \frac{v_{\rm{esc}}}{3}.
\end{equation}
For a typical TDE with a MS star of mass of $0.5\,M_{\odot}$ (with escape velocity of $\sim600\,$km/s) and a BH of mass of $20\,M_{\odot}$, we obtain $v_{\rm{kick}} \simeq 10\,$km/s. These kicks are low compared to both typical cluster escape velocities ($\sim 50-100\,\rm{km\,s}^{-1}$) as well as typical dynamical recoil kicks attained from small-$N$ BH resonant encounters. Thus, these TDE kicks are unlikely to affect the overall BH dynamics in a significant way.

Accounting for these kicks as well as potential BH spin-up as a result of these TDEs is beyond the present study. Follow-up analyses may explore the potential impact of these effects upon the evolution of the BH populations in clusters in more detail.

\subsection{Neutron star TDEs}
\label{sec:NSTDE}

As discussed briefly in Section \ref{sec:method}, \texttt{CMC} records close encounters of neutron stars (NSs) and stars in a manner similar to the BH--star encounters. These NS--star interactions may also lead to disruption events, where the NS acts as the disrupting object. Such events were considered in, e.g., \citet{Hansen1998} and \citet{Perets2016}. In particular, \citet{Hansen1998} argued that accretion onto the NS during such events may trigger collapse to a BH. In total, we identify 28 NS--MS TDEs events in our models 1-24. Roughly, this translates to a galactic rate of $\sim 10^{-8}\,\rm{yr}^{-1}$ per Milky Way-like galaxy, two orders of magnitude lower than the rate predicted for BH--MS TDEs. Thus we conclude that NS--MS TDEs do not occur at an astrophysically interesting rate in this set of cluster models.

As discussed in \citep[e.g.,][]{Ye2018,Belczsynski2018}, the dynamical interaction rate for NSs in a GC is closely related to the cluster's BH population. For clusters with BH populations sufficiently large to dynamically affect the cluster through ``BH heating," the NS population will be relegated to the outer parts of the cluster where densities are relatively low. Only when the BH population has been sufficiently depleted will the NSs see a significant boost in their encounter rate, but even then the encounter rate for NSs is limited by the fact that, on average, NSs have masses of the same order as other populations, in particular white dwarfs and MS binaries. As a result, dynamical interactions involving NSs are never as frequent as those involving BHs in typical GCs, explaining the relatively low rate of NS--MS collisions compared to BH--MS TDEs. However, in a core-collapsed cluster with relatively higher stellar densities and $\sim0$ BHs \citep[see, e.g.,][]{Kremer2019a}, NS--MS TDEs may become more frequent.

\subsection{Link to $r$-process enrichment?}

Observational evidence of $r$-process enrichment exists in several Milky Way GCs, including M5, M15, M92, and NGC 3201 \citep[e.g.,][]{Roederer2011,Bekki2018}. The LIGO/Virgo detection of the binary NS merger GW170817 \citep{Abbott2017} and the follow-up electromagnetic observations showed that NS mergers produce large amounts of $r$-process elements \citep{Kasen2017}. However, explaining the observed $r$-process abundance in GCs with NS merger events proves to be difficult. Recent observational evidence shows that Milky Way GCs exhibit multiple stellar populations which are formed over a series of star formation episodes that can span tens to even hundreds of Myrs \citep[for a recent review on the formation of multiple populations in GCs see, e.g.,][]{Gratton2012}. For a particular binary NS merger event (or series of events) to enrich a GC's stars with $r$-process material, the event(s) must occur while star formation is still taking place (see, e.g., \citet{Bekki2018,Zevin2019} for further discussion).

Merging binary NSs are expected to form in GCs through two mechanisms: binary evolution of primordial binaries \citep[e.g.,][]{Ivanova2003,Dominik2012,Tauris2017} and dynamical formation of NSs at late times \citep[e.g.,][]{Ye2018}. In the first scenario, where the NS components typically form through iron core-collapse supernovae \citep[which are expected to lead to large natal kicks;][]{Hobbs2005}, it is not straightforward to produce binary NSs that remain bound to the cluster or have merger times sufficiently low such that the merger takes place within the cluster environment during GC star formation (\citet{Zevin2019}; also see \citet{Safa2018} for related discussion in the context of ultrafaint dwarf galaxies). In the second scenario, where binary NSs are formed dynamically through exchange encounters, one must wait for the NSs to mass-segregate to the cluster core, which takes $\sim\,$Gyrs, due to the relatively low NS masses \citep[see, e.g.,][]{Ye2018,Zevin2019}. Thus, by the time binary NSs have begun to form dynamically and subsequently merge, any star formation episodes will have almost certainly ceased. In this case, any $r$-process material produced in these late-time NS merger events would be unable to enrich the stellar population.

We propose that if accretion onto a BH arising from the BH--MS disruptions of this study leads to the production of neutron-rich material, these events may provide a way to enrich GCs with $r$-process elements at early times. As shown in \citet{Kremer2018b}, the cluster NGC 3201, one of the MW clusters in which $r$-process enrichment is observed, is consistent with hosting a large population of BHs at present. In particular, our models 7 and 19 have final masses, metallicities, and BH numbers consistent with both observed and theoretical predictions for NGC 3201. In these models, we identify 28 and 35 BH--MS collisions, respectively. Of these, 6 and 2, occur within the first 100 Myr of cluster evolution. (We adopt 100 Myr as an approximate time duration for star formation episodes. The exact duration may be much shorter or longer. Again, see \citet{Gratton2012} for further detail).  Of course, more careful consideration must be taken to determine whether a single BH--MS disruption could produce sufficient (or indeed, any) $r$-process material to explain the observed enrichment. In particular, it is unclear whether the disk expected to form during such an event will reach high enough densities to produce $r$-process elements. We simply note that on the basis of event rates alone the BH--MS disruptions could in principle serve as a possible mechanism. We also note that other mechanisms such as the collapsar model discussed in \citep{Siegel2018} may serve as viable alternatives for $r$-process production in GCs. A complete exploration of $r$-process production in GCs is beyond the scope of this paper; see \citet{Zevin2019} for a more detailed discussion on the topic.

\subsection{Directions for future work}
\label{sec:future}

We note that there are several complexities associated with BH--star TDEs not captured in this analysis. For example, how do effects of metallicity and/or stellar age (and therefore density profile of the star) alter the outcome of these disruption events? Also, are there differences expected if the disrupted star is itself a collision product (i.e., a blue straggler, as discussed briefly in Section \ref{sec:results})? Furthermore, under what circumstances is a relativistic jet expected to be launched? Additionally, recent work \citep[e.g.,][]{Samsing2017, Samsing2018e} has shown that the inclusion of tidal coupling within small-$N$ dynamical encounters may have a significant effect on tidal disruptions. Many of these complexities will require detailed 3-D (magneto)hydrodynamic calculations of these encounters. Our first exploration of some of these complexities will be presented in a forthcoming paper \citep{Fixelle2019}. %We hope to explore these effects within the scope of our full Monte Carlo code in a future study.

The distinction between direct, physical collisions where a BH passes within the radius of a star and more distant encounters near the tidal disruption boundary like those considered in Section \ref{sec:electromagnetic} is likely important when considering the electromagnetic signature of these events. For instance, a direct collision may lead to prompt accretion onto the BH which may make the effects of feedback critical. Here, we have remained agnostic to these differences and simply included the physical collision and more distant tidal disruption limits to bracket the expected range of close encounters between BHs and stars leading to TDEs. More careful treatment of the tidal interactions, especially in the context of binary-mediated resonant encounters, is necessary to explore outcomes of these events in greater detail. In a future study, we hope to explore binary-mediated close encounters between BHs and stars using a small $N$ integrator that self-consistently incorporates relevant hydrodynamic effects to examine some of these complexities in more detail \citep[for previous work on the topic, see, for example,][]{Goodman1991}.

We note that the number of single--single TDEs identified in this analysis is roughly comparable to the number of TDEs through binary encounters (binary--single or binary--binary; compare columns 8, 9, and 10 in Table \ref{table:models}).  This is perhaps surprising, given earlier work showing that the number of close encounters can increase substantially for binary-mediated encounters relative to single--single encounters alone \citep[e.g.,][]{Bacon1996, Fregeau2004,Chatterjee2013}. For example, \citet{Chatterjee2013} showed that binary interactions lead to a marked increase in MS--MS collisions, which may lead to the formation of blue stragglers. Several reasons may explain why we do not see a similar result here in the case of BH--MS TDEs. First, as discussed in \citet{Kremer2018a}, interactions involving BHs and stars are sensitive to the details of the BH--star ``mixing zone,'' the properties of which are determined by a complex interaction between the BH population and the rest of the cluster. In a manner similar to its effect on the formation of accreting BH--star binaries, this interaction may limit the number of binary--mediated BH--star TDEs. Second, as seen in Table 1 in \citet{Chatterjee2013}, the models with the most marked increase in binary-mediated close encounters in that analysis are those with initial global binary fractions as high as $27\%$ (and as high as roughly $40\%$ in the core). Here we consider models with lower binary fraction ($10\%$). A more expansive set of cluster models covering a broader range in initial binary fraction are necessary to evaluate the relative contribution of single--single encounters and binary encounters to the overall TDE rate.

In this analysis, we have focused primarily on stellar-mass BHs, however a large body of literature has also explored the possible existence of IMBHs in GCs from both observational \citep[e.g.,][]{Gebhardt2005,Lanzoni2007,Lutzgendorf2011,Feldmeier2013} and theoretical perspectives \citep[e.g.,][]{PortegiesZwart2004,Freitag2006,Giersz2015,Mapelli2016,Antonini2018}. TDEs (particularly TDEs of WDs) by IMBHs have been examined in various recent analyses including \citet{Rosswog2008,Rosswog2009,MacLeod2014,MacLeod2016b} and these events have been proposed as the mechanisms that may have produced a number of observed high-energy events \citep[e.g.,][]{Krolik2011,Jonker2013}. The existence of IMBHs in clusters remains contested. Nonetheless, the role that IMBHs, if present, may play in the production of high-energy transients like TDEs and also GC dynamics more broadly is a rich topic that we hope to explore in more detail within the scope of our Monte Carlo code in a later study.

\bigskip

\acknowledgments
We thank Sasha Tchekhovskoy, Pablo Marchant, Johan Samsing, Michael Zevin, and Enrico Ramirez-Ruiz for useful discussions. We also acknowledge discussions with Tony Piro and Eliot Quataert on wind reprocessing. This work was supported by NASA ATP Grant NNX14AP92G 
and NSF Grant AST-1716762. KK acknowledges support by the National Science Foundation Graduate Research Fellowship Program under Grant No. DGE-1324585. WL is supported by the David and Ellen Lee Fellowship at Caltech.

\bibliographystyle{aasjournal}
\bibliography{mybib}

\begin{thebibliography}{}
\expandafter\ifx\csname natexlab\endcsname\relax\def\natexlab#1{#1}\fi

\bibitem[{{Abadie} {et~al.}(2010){Abadie}, {Abbott}, {Abbott}, {Abernathy},
  {Accadia}, {Acernese}, {Adams}, {Adhikari}, {Ajith}, {Allen}, \&
  et~al.}]{Abadie2010}
{Abadie}, J., {Abbott}, B.~P., {Abbott}, R., {et~al.} 2010, Classical and
  Quantum Gravity, 27, 173001

\bibitem[{{Abbott} {et~al.}(2016{\natexlab{a}}){Abbott}, {Abbott}, {Abbott},
  {Abernathy}, {Acernese}, {Ackley}, {Adams}, {Adams}, {Addesso}, {Adhikari},
  \& et~al.}]{Abbott2016a}
{Abbott}, B.~P., {Abbott}, R., {Abbott}, T.~D., {et~al.} 2016{\natexlab{a}},
  \apjl, 818, L22

\bibitem[{{Abbott} {et~al.}(2016{\natexlab{b}}){Abbott}, {Abbott}, {Abbott},
  {Abernathy}, {Acernese}, {Ackley}, {Adams}, {Adams}, {Addesso}, {Adhikari},
  \& et~al.}]{Abbott2016c}
---. 2016{\natexlab{b}}, Physical Review Letters, 116, 241103

\bibitem[{{Abbott} {et~al.}(2016{\natexlab{c}}){Abbott}, {Abbott}, {Abbott},
  {Abernathy}, {Acernese}, {Ackley}, {Adams}, {Adams}, {Addesso}, {Adhikari},
  \& et~al.}]{Abbott2016b}
---. 2016{\natexlab{c}}, Physical Review Letters, 116, 061102

\bibitem[{{Abbott} {et~al.}(2016{\natexlab{d}}){Abbott}, {Abbott}, {Abbott},
  {Abernathy}, {Acernese}, {Ackley}, {Adams}, {Adams}, {Addesso}, {Adhikari},
  \& et~al.}]{Abbott2016d}
---. 2016{\natexlab{d}}, Physical Review Letters, 116, 221101

\bibitem[{{Abbott} {et~al.}(2017){Abbott}, {Abbott}, {Abbott}, {Acernese},
  {Ackley}, {Adams}, {Adams}, {Addesso}, {Adhikari}, {Adya}, \&
  et~al.}]{Abbott2017}
---. 2017, Physical Review Letters, 118, 221101

\bibitem[{{Abramowicz} \& {Fragile}(2013)}]{2013LRR....16....1A}
{Abramowicz}, M.~A., \& {Fragile}, P.~C. 2013, Living Reviews in Relativity,
  16, 1

\bibitem[{{Alexander} \& {Kumar}(2001)}]{AlexanderKumar2001}
{Alexander}, T., \& {Kumar}, P. 2001, \apj, 549, 948

\bibitem[{{Antonini} {et~al.}(2018){Antonini}, {Gieles}, \&
  {Gualandris}}]{Antonini2018}
{Antonini}, F., {Gieles}, M., \& {Gualandris}, A. 2018, arXiv e-prints,
  arXiv:1811.03640

\bibitem[{{Arca Sedda} {et~al.}(2018){Arca Sedda}, {Askar}, \&
  {Giersz}}]{ArcaSedda2018}
{Arca Sedda}, M., {Askar}, A., \& {Giersz}, M. 2018, \mnras, 479, 4652

\bibitem[{{Askar} {et~al.}(2017){Askar}, {Szkudlarek}, {Gondek-Rosi{\'n}ska},
  {Giersz}, \& {Bulik}}]{Askar2017}
{Askar}, A., {Szkudlarek}, M., {Gondek-Rosi{\'n}ska}, D., {Giersz}, M., \&
  {Bulik}, T. 2017, \mnras, 464, L36

\bibitem[{{Bacon} {et~al.}(1996){Bacon}, {Sigurdsson}, \& {Davies}}]{Bacon1996}
{Bacon}, D., {Sigurdsson}, S., \& {Davies}, M.~B. 1996, \mnras, 281, 830

\bibitem[{{Bae} {et~al.}(2014){Bae}, {Kim}, \& {Lee}}]{Bae2014}
{Bae}, Y.-B., {Kim}, C., \& {Lee}, H.~M. 2014, \mnras, 440, 2714

\bibitem[{{Bahramian} {et~al.}(2017){Bahramian}, {Heinke}, {Tudor},
  {Miller-Jones}, {Bogdanov}, {Maccarone}, {Knigge}, {Sivakoff}, {Chomiuk},
  {Strader}, {Garcia}, \& {Kallman}}]{Bahramian2017}
{Bahramian}, A., {Heinke}, C.~O., {Tudor}, V., {et~al.} 2017, \mnras, 467, 2199

\bibitem[{{Banerjee}(2017)}]{Banerjee2017}
{Banerjee}, S. 2017, \mnras, 467, 524

\bibitem[{Banerjee {et~al.}(2010)Banerjee, Baumgardt, \& Kroupa}]{Banerjee2010}
Banerjee, S., Baumgardt, H., \& Kroupa, P. 2010, Mon.~Not.~R.~Astron.~Soc, 402,
  371

\bibitem[{{Begelman}(1979)}]{1979MNRAS.187..237B}
{Begelman}, M.~C. 1979, \mnras, 187, 237

\bibitem[{{Begelman}(2012)}]{2012MNRAS.420.2912B}
---. 2012, \mnras, 420, 2912

\bibitem[{{Bekki}(2018)}]{Bekki2018}
{Bekki}, K. 2018, arXiv e-prints, arXiv:1807.02309

\bibitem[{{Belczynski} {et~al.}(2002){Belczynski}, {Kalogera}, \&
  {Bulik}}]{Belczynski2002}
{Belczynski}, K., {Kalogera}, V., \& {Bulik}, T. 2002, \apj, 572, 407

\bibitem[{Belczynski {et~al.}(2016)Belczynski, Heger, Gladysz, Ruiter, Woosley,
  Wiktorowicz, Chen, Bulik, O’Shaughnessy, Holz, Fryer, \&
  Berti}]{Belczynski2016b}
Belczynski, K., Heger, A., Gladysz, W., {et~al.} 2016, Astronomy {\&}
  Astrophysics, 594, A97

\bibitem[{{Belczynski} {et~al.}(2018){Belczynski}, {Askar}, {Arca-Sedda},
  {Chruslinska}, {Donnari}, {Giersz}, {Benacquista}, {Spurzem}, {Jin}, \&
  {Wiktorowicz}}]{Belczsynski2018}
{Belczynski}, K., {Askar}, A., {Arca-Sedda}, M., {et~al.} 2018, \aap, 615, A91

\bibitem[{{Blaes}(2014)}]{2014SSRv..183...21B}
{Blaes}, O. 2014, \ssr, 183, 21

\bibitem[{{Blaes} {et~al.}(2011){Blaes}, {Krolik}, {Hirose}, \&
  {Shabaltas}}]{2011ApJ...733..110B}
{Blaes}, O., {Krolik}, J.~H., {Hirose}, S., \& {Shabaltas}, N. 2011, \apj, 733,
  110

\bibitem[{{Blandford} \& {Begelman}(1999)}]{1999MNRAS.303L...1B}
{Blandford}, R.~D., \& {Begelman}, M.~C. 1999, \mnras, 303, L1

\bibitem[{{Blandford} \& {Begelman}(2004)}]{2004MNRAS.349...68B}
---. 2004, \mnras, 349, 68

\bibitem[{{Campanelli} {et~al.}(2007){Campanelli}, {Lousto}, {Zlochower}, \&
  {Merritt}}]{Campanelli2007}
{Campanelli}, M., {Lousto}, C.~O., {Zlochower}, Y., \& {Merritt}, D. 2007,
  Physical Review Letters, 98, 231102

\bibitem[{Chatterjee {et~al.}(2010)Chatterjee, Fregeau, Umbreit, \&
  Rasio}]{Chatterjee2010}
Chatterjee, S., Fregeau, J.~M., Umbreit, S., \& Rasio, F.~A. 2010, The
  Astrophysical Journal, 719, 915

\bibitem[{{Chatterjee} {et~al.}(2013){Chatterjee}, {Rasio}, {Sills}, \&
  {Glebbeek}}]{Chatterjee2013b}
{Chatterjee}, S., {Rasio}, F.~A., {Sills}, A., \& {Glebbeek}, E. 2013, \apj,
  777, 106

\bibitem[{Chatterjee {et~al.}(2013)Chatterjee, Umbreit, Fregeau, \&
  Rasio}]{Chatterjee2013}
Chatterjee, S., Umbreit, S., Fregeau, J.~M., \& Rasio, F.~A. 2013, Monthly
  Notices of the Royal Astronomical Society, 429, 2881

\bibitem[{Clark(1975)}]{Clark1975}
Clark, G. 1975, \apj, 199, L143

\bibitem[{{Curd} \& {Narayan}(2019)}]{2019MNRAS.483..565C}
{Curd}, B., \& {Narayan}, R. 2019, \mnras, 483, 565

\bibitem[{{Dai} {et~al.}(2018){Dai}, {McKinney}, {Roth}, {Ramirez-Ruiz}, \&
  {Miller}}]{Dai2018}
{Dai}, L., {McKinney}, J.~C., {Roth}, N., {Ramirez-Ruiz}, E., \& {Miller},
  M.~C. 2018, \apjl, 859, L20

\bibitem[{{Dominik} {et~al.}(2012){Dominik}, {Belczynski}, {Fryer}, {Holz},
  {Berti}, {Bulik}, {Mandel}, \& {O'Shaughnessy}}]{Dominik2012}
{Dominik}, M., {Belczynski}, K., {Fryer}, C., {et~al.} 2012, \apj, 759, 52

\bibitem[{{El-Badry} {et~al.}(2018){El-Badry}, {Quataert}, {Weisz}, {Choksi},
  \& {Boylan-Kolchin}}]{El-Badry2018}
{El-Badry}, K., {Quataert}, E., {Weisz}, D.~R., {Choksi}, N., \&
  {Boylan-Kolchin}, M. 2018, ArXiv e-prints, arXiv:1805.03652

\bibitem[{{Fabian} {et~al.}(1975){Fabian}, {Pringle}, \& {Rees}}]{Fabian1975}
{Fabian}, A.~C., {Pringle}, J.~E., \& {Rees}, M.~J. 1975, \mnras, 172, 15p

\bibitem[{{Feldmeier} {et~al.}(2013){Feldmeier}, {L{\"u}tzgendorf}, {Neumayer},
  {Kissler-Patig}, {Gebhardt}, {Baumgardt}, {Noyola}, {de Zeeuw}, \&
  {Jalali}}]{Feldmeier2013}
{Feldmeier}, A., {L{\"u}tzgendorf}, N., {Neumayer}, N., {et~al.} 2013, \aap,
  554, A63

\bibitem[{{Ferraro} {et~al.}(1995){Ferraro}, {Fusi Pecci}, \&
  {Bellazzini}}]{Ferraro1995}
{Ferraro}, F.~R., {Fusi Pecci}, F., \& {Bellazzini}, M. 1995, \aap, 294, 80

\bibitem[{{Fixelle} {et~al.}(2019){Fixelle}, {Kremer}, {Lombardi}, \&
  {Rasio}}]{Fixelle2019}
{Fixelle}, J., {Kremer}, K., {Lombardi}, J., \& {Rasio}, F. 2019

\bibitem[{{Fragione} \& {Kocsis}(2018)}]{Fragione2018b}
{Fragione}, G., \& {Kocsis}, B. 2018, Physical Review Letters, 121, 161103

\bibitem[{{Fragione} {et~al.}(2019){Fragione}, {Leigh}, {Perna}, \&
  {Kocsis}}]{Fragione2019}
{Fragione}, G., {Leigh}, N., {Perna}, R., \& {Kocsis}, B. 2019, arXiv e-prints,
  arXiv:1905.09471

\bibitem[{{Fragione} {et~al.}(2018){Fragione}, {Pavl{\'{\i}}k}, \&
  {Banerjee}}]{Fragione2018a}
{Fragione}, G., {Pavl{\'{\i}}k}, V., \& {Banerjee}, S. 2018, \mnras, 480, 4955

\bibitem[{{Fregeau} {et~al.}(2004){Fregeau}, {Cheung}, {Portegies Zwart}, \&
  {Rasio}}]{Fregeau2004}
{Fregeau}, J.~M., {Cheung}, P., {Portegies Zwart}, S.~F., \& {Rasio}, F.~A.
  2004, \mnras, 352, 1

\bibitem[{Fregeau {et~al.}(2003)Fregeau, Gurkan, Joshi, \& Rasio}]{Fregeau2003}
Fregeau, J.~M., Gurkan, M.~A., Joshi, K.~J., \& Rasio, F.~A. 2003, arXiv,
  astro-ph, 772

\bibitem[{{Fregeau} \& {Rasio}(2007)}]{Fregeau2007}
{Fregeau}, J.~M., \& {Rasio}, F.~A. 2007, \apj, 658, 1047

\bibitem[{{Freitag} \& {Benz}(2002)}]{Freitag2002}
{Freitag}, M., \& {Benz}, W. 2002, \aap, 394, 345

\bibitem[{{Freitag} {et~al.}(2006){Freitag}, {G{\"u}rkan}, \&
  {Rasio}}]{Freitag2006}
{Freitag}, M., {G{\"u}rkan}, M.~A., \& {Rasio}, F.~A. 2006, \mnras, 368, 141

\bibitem[{Fryer \& Kalogera(2001)}]{Fryer2001}
Fryer, C.~L., \& Kalogera, V. 2001, The Astrophysical Journal, 554, 548

\bibitem[{{Fryer} \& {Woosley}(1998)}]{FryerWoosley1998}
{Fryer}, C.~L., \& {Woosley}, S.~E. 1998, \apjl, 502, L9

\bibitem[{{Gebhardt} {et~al.}(2005){Gebhardt}, {Rich}, \& {Ho}}]{Gebhardt2005}
{Gebhardt}, K., {Rich}, R.~M., \& {Ho}, L.~C. 2005, \apj, 634, 1093

\bibitem[{{Geller} \& {Mathieu}(2011)}]{Geller2011}
{Geller}, A.~M., \& {Mathieu}, R.~D. 2011, \nat, 478, 356

\bibitem[{{Gendre} {et~al.}(2013){Gendre}, {Stratta}, {Atteia}, {Basa},
  {Bo{\"e}r}, {Coward}, {Cutini}, {D'Elia}, {Howell}, {Klotz}, \&
  {Piro}}]{2013ApJ...766...30G}
{Gendre}, B., {Stratta}, G., {Atteia}, J.~L., {et~al.} 2013, \apj, 766, 30

\bibitem[{{Giersz}(2001)}]{Giersz2001}
{Giersz}, M. 2001, \mnras, 324, 218

\bibitem[{{Giersz} {et~al.}(2015){Giersz}, {Leigh}, {Hypki}, {L{\"u}tzgendorf},
  \& {Askar}}]{Giersz2015}
{Giersz}, M., {Leigh}, N., {Hypki}, A., {L{\"u}tzgendorf}, N., \& {Askar}, A.
  2015, \mnras, 454, 3150

\bibitem[{{Giesers} {et~al.}(2018){Giesers}, {Dreizler}, {Husser}, {Kamann},
  {Anglada Escud{\'e}}, {Brinchmann}, {Carollo}, {Roth}, {Weilbacher}, \&
  {Wisotzki}}]{Giesers2018}
{Giesers}, B., {Dreizler}, S., {Husser}, T.-O., {et~al.} 2018, \mnras, 475, L15

\bibitem[{{Giesler} {et~al.}(2018){Giesler}, {Clausen}, \& {Ott}}]{Giesler2018}
{Giesler}, M., {Clausen}, D., \& {Ott}, C.~D. 2018, \mnras, 477, 1853

\bibitem[{{Goodman} \& {Hernquist}(1991)}]{Goodman1991}
{Goodman}, J., \& {Hernquist}, L. 1991, \apj, 378, 637

\bibitem[{{Goswami} {et~al.}(2012){Goswami}, {Umbreit}, {Bierbaum}, \&
  {Rasio}}]{Goswami2012}
{Goswami}, S., {Umbreit}, S., {Bierbaum}, M., \& {Rasio}, F.~A. 2012, \apj,
  752, 43

\bibitem[{{Gratton} {et~al.}(2012){Gratton}, {Carretta}, \&
  {Bragaglia}}]{Gratton2012}
{Gratton}, R.~G., {Carretta}, E., \& {Bragaglia}, A. 2012, \aapr, 20, 50

\bibitem[{{Greiner} {et~al.}(2015){Greiner}, {Mazzali}, {Kann}, {Kr{\"u}hler},
  {Pian}, {Prentice}, {Olivares E.}, {Rossi}, {Klose}, {Taubenberger}, {Knust},
  {Afonso}, {Ashall}, {Bolmer}, {Delvaux}, {Diehl}, {Elliott}, {Filgas},
  {Fynbo}, {Graham}, {Guelbenzu}, {Kobayashi}, {Leloudas}, {Savaglio},
  {Schady}, {Schmidl}, {Schweyer}, {Sudilovsky}, {Tanga}, {Updike}, {van
  Eerten}, \& {Varela}}]{2015Natur.523..189G}
{Greiner}, J., {Mazzali}, P.~A., {Kann}, D.~A., {et~al.} 2015, \nat, 523, 189

\bibitem[{{Guillochon} \& {Ramirez-Ruiz}(2013)}]{2013ApJ...767...25G}
{Guillochon}, J., \& {Ramirez-Ruiz}, E. 2013, \apj, 767, 25

\bibitem[{{Hansen} \& {Murali}(1998)}]{Hansen1998}
{Hansen}, B.~M.~S., \& {Murali}, C. 1998, \apjl, 505, L15

\bibitem[{{Heggie} \& {Hut}(2003)}]{HeggieHut2003}
{Heggie}, D., \& {Hut}, P. 2003, {The Gravitational Million-Body Problem: A
  Multidisciplinary Approach to Star Cluster Dynamics}

\bibitem[{{Heinke} {et~al.}(2005){Heinke}, {Grindlay}, {Edmonds}, {Cohn},
  {Lugger}, {Camilo}, {Bogdanov}, \& {Freire}}]{Heinke2005}
{Heinke}, C.~O., {Grindlay}, J.~E., {Edmonds}, P.~D., {et~al.} 2005, \apj, 625,
  796

\bibitem[{Henon(1971)}]{Henon1971a}
Henon, M. 1971, Astrophysics and Space Science, 13, 284

\bibitem[{H{\'{e}}non(1971)}]{Henon1971b}
H{\'{e}}non, M. 1971, Astrophysics and Space Science, 14, 151

\bibitem[{{Hills}(1976)}]{Hills1976}
{Hills}, J.~G. 1976, \mnras, 175, 1P

\bibitem[{Hobbs {et~al.}(2005)Hobbs, Lorimer, Lyne, \& Kramer}]{Hobbs2005}
Hobbs, G., Lorimer, D.~R., Lyne, A.~G., \& Kramer, M. 2005, Monthly Notices of
  the Royal Astronomical Society, 360, 974

\bibitem[{{Hong} {et~al.}(2018){Hong}, {Vesperini}, {Askar}, {Giersz},
  {Szkudlarek}, \& {Bulik}}]{Hong2018}
{Hong}, J., {Vesperini}, E., {Askar}, A., {et~al.} 2018, \mnras, 480, 5645

\bibitem[{{Hurley} {et~al.}(2000){Hurley}, {Pols}, \& {Tout}}]{Hurley2000}
{Hurley}, J.~R., {Pols}, O.~R., \& {Tout}, C.~A. 2000, \mnras, 315, 543

\bibitem[{{Hurley} {et~al.}(2002){Hurley}, {Tout}, \& {Pols}}]{Hurley2002}
{Hurley}, J.~R., {Tout}, C.~A., \& {Pols}, O.~R. 2002, \mnras, 329, 897

\bibitem[{{Ivanova}(2013)}]{Ivanova2013}
{Ivanova}, N. 2013, \memsai, 84, 123

\bibitem[{{Ivanova} {et~al.}(2003){Ivanova}, {Belczynski}, {Kalogera}, {Rasio},
  \& {Taam}}]{Ivanova2003}
{Ivanova}, N., {Belczynski}, K., {Kalogera}, V., {Rasio}, F.~A., \& {Taam},
  R.~E. 2003, \apj, 592, 475

\bibitem[{{Ivanova} {et~al.}(2010){Ivanova}, {Chaichenets}, {Fregeau},
  {Heinke}, {Lombardi}, \& {Woods}}]{Ivanova2010}
{Ivanova}, N., {Chaichenets}, S., {Fregeau}, J., {et~al.} 2010, \apj, 717, 948

\bibitem[{{Ivanova} {et~al.}(2017){Ivanova}, {da Rocha}, {Van}, \&
  {Nandez}}]{Ivanova2017}
{Ivanova}, N., {da Rocha}, C.~A., {Van}, K.~X., \& {Nandez}, J.~L.~A. 2017,
  \apjl, 843, L30

\bibitem[{{Ivanova} {et~al.}(2008){Ivanova}, {Heinke}, {Rasio}, {Belczynski},
  \& {Fregeau}}]{Ivanova2008}
{Ivanova}, N., {Heinke}, C.~O., {Rasio}, F.~A., {Belczynski}, K., \& {Fregeau},
  J.~M. 2008, \mnras, 386, 553

\bibitem[{{Ivanova} {et~al.}(2005){Ivanova}, {Rasio}, {Lombardi}, {Dooley}, \&
  {Proulx}}]{Ivanova2005}
{Ivanova}, N., {Rasio}, F.~A., {Lombardi}, Jr., J.~C., {Dooley}, K.~L., \&
  {Proulx}, Z.~F. 2005, \apjl, 621, L109

\bibitem[{{Jiang} {et~al.}(2017){Jiang}, {Stone}, \&
  {Davis}}]{2017arXiv170902845J}
{Jiang}, Y.-F., {Stone}, J., \& {Davis}, S.~W. 2017, arXiv e-prints,
  arXiv:1709.02845

\bibitem[{{Jonker} {et~al.}(2013){Jonker}, {Glennie}, {Heida}, {Maccarone},
  {Hodgkin}, {Nelemans}, {Miller-Jones}, {Torres}, \& {Fender}}]{Jonker2013}
{Jonker}, P.~G., {Glennie}, A., {Heida}, M., {et~al.} 2013, \apj, 779, 14

\bibitem[{Joshi {et~al.}(2001)Joshi, Nave, \& Rasio}]{Joshi2001}
Joshi, K.~J., Nave, C.~P., \& Rasio, F.~A. 2001, The Astrophysical Journal,
  550, 691

\bibitem[{Joshi {et~al.}(2000)Joshi, Rasio, Zwart, \&
  Portegies~Zwart}]{Joshi2000}
Joshi, K.~J., Rasio, F.~A., Zwart, S.~P., \& Portegies~Zwart, S. 2000, The
  Astrophysical Journal, 540, 969

\bibitem[{{Kasen} {et~al.}(2017){Kasen}, {Metzger}, {Barnes}, {Quataert}, \&
  {Ramirez-Ruiz}}]{Kasen2017}
{Kasen}, D., {Metzger}, B., {Barnes}, J., {Quataert}, E., \& {Ramirez-Ruiz}, E.
  2017, \nat, 551, 80

\bibitem[{{Kiel} \& {Hurley}(2009)}]{KielHurley2009}
{Kiel}, P.~D., \& {Hurley}, J.~R. 2009, \mnras, 395, 2326

\bibitem[{{Kiel} {et~al.}(2008){Kiel}, {Hurley}, {Bailes}, \&
  {Murray}}]{Kiel2008}
{Kiel}, P.~D., {Hurley}, J.~R., {Bailes}, M., \& {Murray}, J.~R. 2008, \mnras,
  388, 393

\bibitem[{{Kremer} {et~al.}(2018{\natexlab{a}}){Kremer}, {Chatterjee},
  {Breivik}, {Rodriguez}, {Larson}, \& {Rasio}}]{Kremer2018c}
{Kremer}, K., {Chatterjee}, S., {Breivik}, K., {et~al.} 2018{\natexlab{a}},
  Physical Review Letters, 120, 191103

\bibitem[{{Kremer} {et~al.}(2018{\natexlab{b}}){Kremer}, {Chatterjee},
  {Rodriguez}, \& {Rasio}}]{Kremer2018a}
{Kremer}, K., {Chatterjee}, S., {Rodriguez}, C.~L., \& {Rasio}, F.~A.
  2018{\natexlab{b}}, \apj, 852, 29

\bibitem[{{Kremer} {et~al.}(2019){Kremer}, {Chatterjee}, {Ye}, {Rodriguez}, \&
  {Rasio}}]{Kremer2019a}
{Kremer}, K., {Chatterjee}, S., {Ye}, C.~S., {Rodriguez}, C.~L., \& {Rasio},
  F.~A. 2019, \apj, 871, 38

\bibitem[{{Kremer} {et~al.}(2018{\natexlab{c}}){Kremer}, {Ye}, {Chatterjee},
  {Rodriguez}, \& {Rasio}}]{Kremer2018b}
{Kremer}, K., {Ye}, C.~S., {Chatterjee}, S., {Rodriguez}, C.~L., \& {Rasio},
  F.~A. 2018{\natexlab{c}}, \apjl, 855, L15

\bibitem[{{Krolik}(1984)}]{Krolik1984}
{Krolik}, J.~H. 1984, \apj, 282, 452

\bibitem[{{Krolik} \& {Piran}(2011)}]{Krolik2011}
{Krolik}, J.~H., \& {Piran}, T. 2011, \apj, 743, 134

\bibitem[{{Kroupa}(2001)}]{Kroupa2001}
{Kroupa}, P. 2001, \mnras, 322, 231

\bibitem[{{Kumar} {et~al.}(2008){Kumar}, {Narayan}, \&
  {Johnson}}]{2008MNRAS.388.1729K}
{Kumar}, P., {Narayan}, R., \& {Johnson}, J.~L. 2008, \mnras, 388, 1729

\bibitem[{{Lanzoni} {et~al.}(2007){Lanzoni}, {Dalessandro}, {Ferraro},
  {Miocchi}, {Valenti}, \& {Rood}}]{Lanzoni2007}
{Lanzoni}, B., {Dalessandro}, E., {Ferraro}, F.~R., {et~al.} 2007, \apj, 668,
  L139

\bibitem[{{Leigh} {et~al.}(2013){Leigh}, {Knigge}, {Sills}, {Perets},
  {Sarajedini}, \& {Glebbeek}}]{Leigh2013}
{Leigh}, N., {Knigge}, C., {Sills}, A., {et~al.} 2013, \mnras, 428, 897

\bibitem[{{Levan} {et~al.}(2014){Levan}, {Tanvir}, {Starling}, {Wiersema},
  {Page}, {Perley}, {Schulze}, {Wynn}, {Chornock}, {Hjorth}, {Cenko},
  {Fruchter}, {O'Brien}, {Brown}, {Tunnicliffe}, {Malesani}, {Jakobsson},
  {Watson}, {Berger}, {Bersier}, {Cobb}, {Covino}, {Cucchiara}, {de Ugarte
  Postigo}, {Fox}, {Gal-Yam}, {Goldoni}, {Gorosabel}, {Kaper}, {Kr{\"u}hler},
  {Karjalainen}, {Osborne}, {Pian}, {S{\'a}nchez-Ram{\'\i}rez}, {Schmidt},
  {Skillen}, {Tagliaferri}, {Th{\"o}ne}, {Vaduvescu}, {Wijers}, \&
  {Zauderer}}]{2014ApJ...781...13L}
{Levan}, A.~J., {Tanvir}, N.~R., {Starling}, R.~L.~C., {et~al.} 2014, \apj,
  781, 13

\bibitem[{{Lopez} {et~al.}(2018){Lopez}, {Batta}, {Ramirez-Ruiz}, {Martinez},
  \& {Samsing}}]{Lopez2018}
{Lopez}, Jr., M., {Batta}, A., {Ramirez-Ruiz}, E., {Martinez}, I., \&
  {Samsing}, J. 2018, arXiv e-prints, arXiv:1812.01118

\bibitem[{{Lousto} {et~al.}(2012){Lousto}, {Zlochower}, {Dotti}, \&
  {Volonteri}}]{Lousto2012}
{Lousto}, C.~O., {Zlochower}, Y., {Dotti}, M., \& {Volonteri}, M. 2012, \prd,
  85, 084015

\bibitem[{{L{\"u}tzgendorf} {et~al.}(2011){L{\"u}tzgendorf}, {Kissler-Patig},
  {Noyola}, {Jalali}, {de Zeeuw}, {Gebhardt}, \& {Baumgardt}}]{Lutzgendorf2011}
{L{\"u}tzgendorf}, N., {Kissler-Patig}, M., {Noyola}, E., {et~al.} 2011, \aap,
  533, A36

\bibitem[{{Lynden-Bell} \& {Pringle}(1974)}]{1974MNRAS.168..603L}
{Lynden-Bell}, D., \& {Pringle}, J.~E. 1974, \mnras, 168, 603

\bibitem[{Lyne {et~al.}(1987)Lyne, Brinklow, Middleditch, Kulkarni, Backer, \&
  Clifton}]{Lyne1987}
Lyne, A., Brinklow, A., Middleditch, J., {et~al.} 1987, \nat, 328, 399

\bibitem[{{Maccarone} {et~al.}(2007){Maccarone}, {Kundu}, {Zepf}, \&
  {Rhode}}]{Maccarone2007}
{Maccarone}, T.~J., {Kundu}, A., {Zepf}, S.~E., \& {Rhode}, K.~L. 2007, \nat,
  445, 183

\bibitem[{{Mackey} {et~al.}(2007){Mackey}, {Wilkinson}, {Davies}, \&
  {Gilmore}}]{Mackey2007}
{Mackey}, A.~D., {Wilkinson}, M.~I., {Davies}, M.~B., \& {Gilmore}, G.~F. 2007,
  \mnras, 379, L40

\bibitem[{{Mackey} {et~al.}(2008){Mackey}, {Wilkinson}, {Davies}, \&
  {Gilmore}}]{Mackey2008}
---. 2008, \mnras, 386, 65

\bibitem[{{MacLeod} {et~al.}(2014){MacLeod}, {Goldstein}, {Ramirez-Ruiz},
  {Guillochon}, \& {Samsing}}]{MacLeod2014}
{MacLeod}, M., {Goldstein}, J., {Ramirez-Ruiz}, E., {Guillochon}, J., \&
  {Samsing}, J. 2014, \apj, 794, 9

\bibitem[{{MacLeod} {et~al.}(2016){MacLeod}, {Trenti}, \&
  {Ramirez-Ruiz}}]{MacLeod2016b}
{MacLeod}, M., {Trenti}, M., \& {Ramirez-Ruiz}, E. 2016, \apj, 819, 70

\bibitem[{{Mapelli}(2016)}]{Mapelli2016}
{Mapelli}, M. 2016, \mnras, 459, 3432

\bibitem[{{McKinney} {et~al.}(2015){McKinney}, {Dai}, \&
  {Avara}}]{2015MNRAS.454L...6M}
{McKinney}, J.~C., {Dai}, L., \& {Avara}, M.~J. 2015, \mnras, 454, L6

\bibitem[{{McKinney} {et~al.}(2014){McKinney}, {Tchekhovskoy}, {Sadowski}, \&
  {Narayan}}]{2014MNRAS.441.3177M}
{McKinney}, J.~C., {Tchekhovskoy}, A., {Sadowski}, A., \& {Narayan}, R. 2014,
  \mnras, 441, 3177

\bibitem[{{Merritt} {et~al.}(2004){Merritt}, {Piatek}, {Portegies Zwart}, \&
  {Hemsendorf}}]{Merritt2004}
{Merritt}, D., {Piatek}, S., {Portegies Zwart}, S., \& {Hemsendorf}, M. 2004,
  \apj, 608, L25

\bibitem[{{Moody} \& {Sigurdsson}(2009)}]{Moody2009}
{Moody}, K., \& {Sigurdsson}, S. 2009, \apj, 690, 1370

\bibitem[{Morscher {et~al.}(2015)Morscher, Pattabiraman, Rodriguez, Rasio, \&
  Umbreit}]{Morscher2015}
Morscher, M., Pattabiraman, B., Rodriguez, C., Rasio, F.~A., \& Umbreit, S.
  2015, The Astrophysical Journal, 800, 9

\bibitem[{{Narayan} {et~al.}(2017){Narayan}, {Sa{\`I}{\textsection}dowski}, \&
  {Soria}}]{2017MNRAS.469.2997N}
{Narayan}, R., {Sa{\`I}{\textsection}dowski}, A., \& {Soria}, R. 2017, \mnras,
  469, 2997

\bibitem[{{Narayan} \& {Yi}(1994)}]{1994ApJ...428L..13N}
{Narayan}, R., \& {Yi}, I. 1994, \apjl, 428, L13

\bibitem[{{Ohsuga} \& {Mineshige}(2011)}]{2011ApJ...736....2O}
{Ohsuga}, K., \& {Mineshige}, S. 2011, \apj, 736, 2

\bibitem[{Pattabiraman {et~al.}(2013)Pattabiraman, Umbreit, Liao, Choudhary,
  Kalogera, Memik, \& Rasio}]{Pattabiraman2013}
Pattabiraman, B., Umbreit, S., Liao, W.-k., {et~al.} 2013, The Astrophysical
  Journal Supplement Series, 204, 15

\bibitem[{{Perets} {et~al.}(2016){Perets}, {Li}, {Lombardi}, \&
  {Milcarek}}]{Perets2016}
{Perets}, H.~B., {Li}, Z., {Lombardi}, Jr., J.~C., \& {Milcarek}, Jr., S.~R.
  2016, \apj, 823, 113

\bibitem[{{Peuten} {et~al.}(2016){Peuten}, {Zocchi}, {Gieles}, {Gualandris}, \&
  {H{\'e}nault-Brunet}}]{Peuten2016}
{Peuten}, M., {Zocchi}, A., {Gieles}, M., {Gualandris}, A., \&
  {H{\'e}nault-Brunet}, V. 2016, \mnras, 462, 2333

\bibitem[{{Piotto} {et~al.}(2002){Piotto}, {King}, {Djorgovski}, {Sosin},
  {Zoccali}, {Saviane}, {de Angeli}, {Riello}, {Recio-Blanco}, {Rich},
  {Meylan}, \& {Renzini}}]{Piotto2003}
{Piotto}, G., {King}, I.~R., {Djorgovski}, S.~G., {et~al.} 2002, VizieR Online
  Data Catalog, 339

\bibitem[{{Portegies Zwart} {et~al.}(2004){Portegies Zwart}, {Baumgardt},
  {Hut}, {Makino}, \& {McMillan}}]{PortegiesZwart2004}
{Portegies Zwart}, S.~F., {Baumgardt}, H., {Hut}, P., {Makino}, J., \&
  {McMillan}, S. L.~W. 2004, \nat, 428, 724

\bibitem[{{Portegies Zwart} {et~al.}(2010){Portegies Zwart}, {McMillan}, \&
  {Gieles}}]{PortegiesZwart2010}
{Portegies Zwart}, S.~F., {McMillan}, S.~L.~W., \& {Gieles}, M. 2010, \araa,
  48, 431

\bibitem[{{Ransom}(2008)}]{Ransom2008}
{Ransom}, S.~M. 2008, in IAU Symposium, Vol. 246, Dynamical Evolution of Dense
  Stellar Systems, ed. E.~{Vesperini}, M.~{Giersz}, \& A.~{Sills}, 291--300

\bibitem[{{Rodriguez} {et~al.}(2018{\natexlab{a}}){Rodriguez}, {Amaro-Seoane},
  {Chatterjee}, {Kremer}, {Rasio}, {Samsing}, {Ye}, \&
  {Zevin}}]{Rodriguez2018b}
{Rodriguez}, C.~L., {Amaro-Seoane}, P., {Chatterjee}, S., {et~al.}
  2018{\natexlab{a}}, \prd, 98, 123005

\bibitem[{{Rodriguez} {et~al.}(2018{\natexlab{b}}){Rodriguez}, {Amaro-Seoane},
  {Chatterjee}, \& {Rasio}}]{Rodriguez2018a}
{Rodriguez}, C.~L., {Amaro-Seoane}, P., {Chatterjee}, S., \& {Rasio}, F.~A.
  2018{\natexlab{b}}, Physical Review Letters, 120, 151101

\bibitem[{Rodriguez {et~al.}(2016)Rodriguez, Chatterjee, \&
  Rasio}]{Rodriguez2016a}
Rodriguez, C.~L., Chatterjee, S., \& Rasio, F.~A. 2016, Physical Review D, 93,
  084029

\bibitem[{{Rodriguez} \& {Loeb}(2018)}]{RodriguezLoeb2018}
{Rodriguez}, C.~L., \& {Loeb}, A. 2018, ArXiv e-prints, arXiv:1809.01152

\bibitem[{Rodriguez {et~al.}(2015)Rodriguez, Morscher, Pattabiraman,
  Chatterjee, Haster, \& Rasio}]{Rodriguez2015a}
Rodriguez, C.~L., Morscher, M., Pattabiraman, B., {et~al.} 2015, Physical
  Review Letters, 115, 051101

\bibitem[{{Roederer}(2011)}]{Roederer2011}
{Roederer}, I.~U. 2011, \apjl, 732, L17

\bibitem[{{Rosswog} {et~al.}(2008){Rosswog}, {Ramirez-Ruiz}, \&
  {Hix}}]{Rosswog2008}
{Rosswog}, S., {Ramirez-Ruiz}, E., \& {Hix}, W.~R. 2008, \apj, 679, 1385

\bibitem[{{Rosswog} {et~al.}(2009){Rosswog}, {Ramirez-Ruiz}, \&
  {Hix}}]{Rosswog2009}
---. 2009, \apj, 695, 404

\bibitem[{{Rybicki} \& {Lightman}(1979)}]{1979rpa..book.....R}
{Rybicki}, G.~B., \& {Lightman}, A.~P. 1979, {Radiative processes in
  astrophysics}

\bibitem[{{Safarzadeh} {et~al.}(2018){Safarzadeh}, {Ramirez-Ruiz}, {Andrews},
  {Fragos}, {Macias}, \& {Scannapieco}}]{Safa2018}
{Safarzadeh}, M., {Ramirez-Ruiz}, E., {Andrews}, J.~J., {et~al.} 2018, arXiv
  e-prints, arXiv:1810.04176

\bibitem[{{Samsing} \& {D'Orazio}(2018)}]{Samsing2018a}
{Samsing}, J., \& {D'Orazio}, D.~J. 2018, \mnras, arXiv:1804.06519

\bibitem[{{Samsing} {et~al.}(2018){Samsing}, {Leigh}, \&
  {Trani}}]{Samsing2018e}
{Samsing}, J., {Leigh}, N.~W.~C., \& {Trani}, A.~A. 2018, \mnras, 481, 5436

\bibitem[{{Samsing} {et~al.}(2017){Samsing}, {MacLeod}, \&
  {Ramirez-Ruiz}}]{Samsing2017}
{Samsing}, J., {MacLeod}, M., \& {Ramirez-Ruiz}, E. 2017, \apj, 846, 36

\bibitem[{{Samsing} {et~al.}(2019){Samsing}, {Venumadhav}, {Dai}, {Martinez},
  {Batta}, {Lopez}, {Ramirez-Ruiz}, \& {Kremer}}]{Samsing2019}
{Samsing}, J., {Venumadhav}, T., {Dai}, L., {et~al.} 2019, arXiv e-prints,
  arXiv:1901.02889

\bibitem[{{Sandage}(1953)}]{Sandage1953}
{Sandage}, A.~R. 1953, \aj, 58, 61

\bibitem[{{S{\c a}dowski} \& {Narayan}(2016)}]{2016MNRAS.456.3929S}
{S{\c a}dowski}, A., \& {Narayan}, R. 2016, \mnras, 456, 3929

\bibitem[{{Shakura} \& {Sunyaev}(1973)}]{1973A&A....24..337S}
{Shakura}, N.~I., \& {Sunyaev}, R.~A. 1973, \aap, 24, 337

\bibitem[{{Shen} \& {Matzner}(2014)}]{2014ApJ...784...87S}
{Shen}, R.-F., \& {Matzner}, C.~D. 2014, \apj, 784, 87

\bibitem[{{Shishkovsky} {et~al.}(2018){Shishkovsky}, {Strader}, {Chomiuk},
  {Bahramian}, {Tremou}, {Li}, {Salinas}, {Tudor}, {Miller-Jones}, {Maccarone},
  {Heinke}, \& {Sivakoff}}]{Shishkovsky2018}
{Shishkovsky}, L., {Strader}, J., {Chomiuk}, L., {et~al.} 2018, \apj, 855, 55

\bibitem[{{Siegel} {et~al.}(2018){Siegel}, {Barnes}, \& {Metzger}}]{Siegel2018}
{Siegel}, D.~M., {Barnes}, J., \& {Metzger}, B.~D. 2018, arXiv e-prints,
  arXiv:1810.00098

\bibitem[{{Sigurdsson} \& {Phinney}(1995)}]{Sigurdsson1995}
{Sigurdsson}, S., \& {Phinney}, E.~S. 1995, \apjs, 99, 609

\bibitem[{{Soltan}(1982)}]{1982MNRAS.200..115S}
{Soltan}, A. 1982, \mnras, 200, 115

\bibitem[{{Spera} \& {Mapelli}(2017)}]{Spera2017}
{Spera}, M., \& {Mapelli}, M. 2017, \mnras, 470, 4739

\bibitem[{{Strader} {et~al.}(2012){Strader}, {Chomiuk}, {Maccarone},
  {Miller-Jones}, \& {Seth}}]{Strader2012}
{Strader}, J., {Chomiuk}, L., {Maccarone}, T.~J., {Miller-Jones}, J.~C.~A., \&
  {Seth}, A.~C. 2012, \nat, 490, 71

\bibitem[{{Strubbe} \& {Quataert}(2009)}]{2009MNRAS.400.2070S}
{Strubbe}, L.~E., \& {Quataert}, E. 2009, \mnras, 400, 2070

\bibitem[{{Tauris} {et~al.}(2017){Tauris}, {Kramer}, {Freire}, {Wex}, {Janka},
  {Langer}, {Podsiadlowski}, {Bozzo}, {Chaty}, {Kruckow}, {van den Heuvel},
  {Antoniadis}, {Breton}, \& {Champion}}]{Tauris2017}
{Tauris}, T.~M., {Kramer}, M., {Freire}, P.~C.~C., {et~al.} 2017, \apj, 846,
  170

\bibitem[{{Tchekhovskoy} {et~al.}(2014){Tchekhovskoy}, {Metzger}, {Giannios},
  \& {Kelley}}]{2014MNRAS.437.2744T}
{Tchekhovskoy}, A., {Metzger}, B.~D., {Giannios}, D., \& {Kelley}, L.~Z. 2014,
  \mnras, 437, 2744

\bibitem[{{Tchekhovskoy} {et~al.}(2011){Tchekhovskoy}, {Narayan}, \&
  {McKinney}}]{2011MNRAS.418L..79T}
{Tchekhovskoy}, A., {Narayan}, R., \& {McKinney}, J.~C. 2011, \mnras, 418, L79

\bibitem[{{The LIGO Scientific Collaboration} {et~al.}(2018){The LIGO
  Scientific Collaboration}, {the Virgo Collaboration}, {Abbott}, {Abbott},
  {Abbott}, {Abraham}, {Acernese}, {Ackley}, {Adams}, {Adhikari}, \&
  et~al.}]{Abbott2018b}
{The LIGO Scientific Collaboration}, {the Virgo Collaboration}, {Abbott},
  B.~P., {et~al.} 2018, arXiv e-prints, arXiv:1811.12907

\bibitem[{{Verbunt} {et~al.}(1984){Verbunt}, {van Paradijs}, \&
  {Elson}}]{Verbunt1984}
{Verbunt}, F., {van Paradijs}, J., \& {Elson}, R. 1984, \mnras, 210, 899

\bibitem[{{Wang} {et~al.}(2016){Wang}, {Spurzem}, {Aarseth}, {Giersz}, {Askar},
  {Berczik}, {Naab}, {Schadow}, \& {Kouwenhoven}}]{Wang2016}
{Wang}, L., {Spurzem}, R., {Aarseth}, S., {et~al.} 2016, \mnras, 458, 1450

\bibitem[{{Woosley}(2016)}]{Woosley2016}
{Woosley}, S.~E. 2016, \apjl, 824, L10

\bibitem[{{Ye} {et~al.}(2019){Ye}, {Kremer}, {Chatterjee}, {Rodriguez}, \&
  {Rasio}}]{Ye2018}
{Ye}, C.~S., {Kremer}, K., {Chatterjee}, S., {Rodriguez}, C.~L., \& {Rasio},
  F.~A. 2019, arXiv e-prints, arXiv:1902.05963

\bibitem[{{Yuan} {et~al.}(2012){Yuan}, {Wu}, \& {Bu}}]{2012ApJ...761..129Y}
{Yuan}, F., {Wu}, M., \& {Bu}, D. 2012, \apj, 761, 129

\bibitem[{{Zevin} {et~al.}(2019){Zevin}, {Kremer}, \& et~al.}]{Zevin2019}
{Zevin}, M., {Kremer}, K., \& et~al. 2019

\bibitem[{{Zhang} \& {Fryer}(2001)}]{ZhangFryer2001}
{Zhang}, W., \& {Fryer}, C.~L. 2001, \apj, 550, 357

\bibitem[{{Ziosi} {et~al.}(2014){Ziosi}, {Mapelli}, {Branchesi}, \&
  {Tormen}}]{Ziosi2014}
{Ziosi}, B.~M., {Mapelli}, M., {Branchesi}, M., \& {Tormen}, G. 2014, \mnras,
  441, 3703

\bibitem[{{Zocchi} {et~al.}(2019){Zocchi}, {Gieles}, \&
  {H{\'e}nault-Brunet}}]{Zocchi2019}
{Zocchi}, A., {Gieles}, M., \& {H{\'e}nault-Brunet}, V. 2019, \mnras, 482, 4713

\end{thebibliography}

\listofchanges

\end{document}